\documentclass[10pt,journal,twocolumn,final]{IEEEtran}

\usepackage{eqnarray,amssymb,amsmath,amsthm,mathtools,amsfonts,etoolbox}
\usepackage{mathrsfs}
\usepackage[utf8]{inputenc}
\usepackage[T1]{fontenc}
\usepackage{graphicx}
\usepackage{subfigure}
\usepackage{epstopdf}
\usepackage{color}
\usepackage{cite}
\usepackage{bbm}
\usepackage{algorithm}
\usepackage{algpseudocode}

\makeatletter
\renewenvironment{cases}[1][\lbrace]{%
  \def\@ldelim{#1}
  \matrix@check\cases\env@cases
}{%
  \endarray\right.%
}
\patchcmd{\env@cases}{\lbrace}{\@ldelim}{}{}
\makeatother


\begin{document}
\title{Bayesian Inverse Contextual Reasoning for Heterogeneous Semantics-Native Communication}

\author{\IEEEauthorblockN{Hyowoon Seo, \IEEEmembership{Member, IEEE}, Yoonseong Kang, \IEEEmembership{Student Member, IEEE}, Mehdi Bennis, \IEEEmembership{Fellow, IEEE}, and Wan Choi, \IEEEmembership{Fellow, IEEE}}
\thanks{H. Seo is with the Department of Electronics and Communications, Kwangwoon University, Seoul, Korea (e-mail: hyowoonseo@kw.ac.kr)}
\thanks{Y. Kang is with the School of Electrical Engineering, Korea Advanced Institute of Science and Technology (KAIST), Daejeon 34141, Korea (e-mail: yoonseongkang@kaist.ac.kr).}
\thanks{M. Bennis is with the Centre for Wireless Communications, University of Oulu, 90014 Oulu, Finland (e-mail: mehdi.bennis@oulu.fi).}
\thanks{W. Choi is with the Institute of New Media and Communications and the Department of Electrical and Computer Engineering, Seoul National University (SNU), Seoul 08826, Korea (e-mail: wanchoi@snu.ac.kr).}
\thanks{(\emph{Corresponding author: Wan Choi})}
}

\maketitle

\begin{abstract}
This work deals with the heterogeneous semantic-native communication  (SNC) problem. When agents do not share the same communication context, the effectiveness of contextual reasoning (CR) is compromised calling for agents to infer other agents’ context. This article proposes a novel framework for solving the inverse problem of CR in SNC using two Bayesian inference methods, namely: Bayesian inverse CR (iCR) and Bayesian inverse linearized CR (iLCR).  The first proposed Bayesian iCR method utilizes Markov Chain Monte Carlo (MCMC) sampling to infer the agent’s context while being computationally expensive. To address this issue, a Bayesian iLCR method is leveraged which obtains a linearized CR (LCR) model by training a linear neural network. Experimental results show that the Bayesian iLCR method requires less computation and achieves higher inference accuracy compared to Bayesian iCR. Additionally, heterogeneous SNC based on the context obtained through the Bayesian iLCR method shows better communication effectiveness than that of Bayesian iCR. Overall, this work provides valuable insights and methods to improve the effectiveness of SNC in situations where agents have different contexts.
\end{abstract}



\section{Introduction}\label{sec:introduction}

Recent advancements in machine learning (ML) have ushered in a new communication paradigm known as semantic communication \cite{Seo2023,Popovski2019,Kountouris2020,Maatouk2020,Qin2022,Luo2022,Xie2021a,Xie2021b}. Semantic communication shows great promise in breaking through the performance bottleneck of existing communication technologies and is envisioned as a key enabler for the sixth-generation (6G) communication systems. The fundamental difference between traditional and semantic communication systems lies in their respective objectives. Traditional communication systems (also known as Shannon's level A) are designed to deliver data accurately, whereas semantic communication (also referred to as Shannon's levels B and C) aims to convey the meaning of data that is used to effectively solve a task. In semantics communication, the effectiveness of the communicating agents' task performance is measured to ensure that data semantics is correctly conveyed. As a result, the primary goal of semantic communication is to maximize communication effectiveness in terms of task performance, while maintaining or improving efficiency \cite{Shannon1964}. 
In recent years, multiple studies have focused on addressing the semantic-effectiveness problem, as evidenced by several research works \cite{Popovski2019,Kountouris2020,Maatouk2020,Wang2021,Qin2022,Luo2022,Xie2021a,Xie2021b,Seo2023,Farshbafan2023}. These works approach the problem from different perspectives but share a common goal of seeking ways to extract meaningful information from data to enhance communication effectiveness and efficiency. The first such study, conducted by Seo et al. \cite{Seo2023}, asserts that data semantics is closely tied to the context within which communicating agents operate. Specifically, the study proposed a stochastic semantic communication model, called \emph{semantics-native communication (SNC)}, which utilizes \emph{contextual reasoning (CR)} inspired by human communication to enhance communication performance. The CR strategy allows communicating agents to consider the context of the communication task, which leads to a better understanding of the meaning of transmitted data. Remarkably, the study demonstrated that the proposed SNC model significantly improves communication effectiveness and efficiency when both communicating agents share a communication context or have knowledge of each other's communication context. The findings suggest that communication systems that harness communication context can potentially enhance communication performance and effectiveness.

\begin{figure}[!t]
	\centering
	\includegraphics[width=0.485\textwidth]{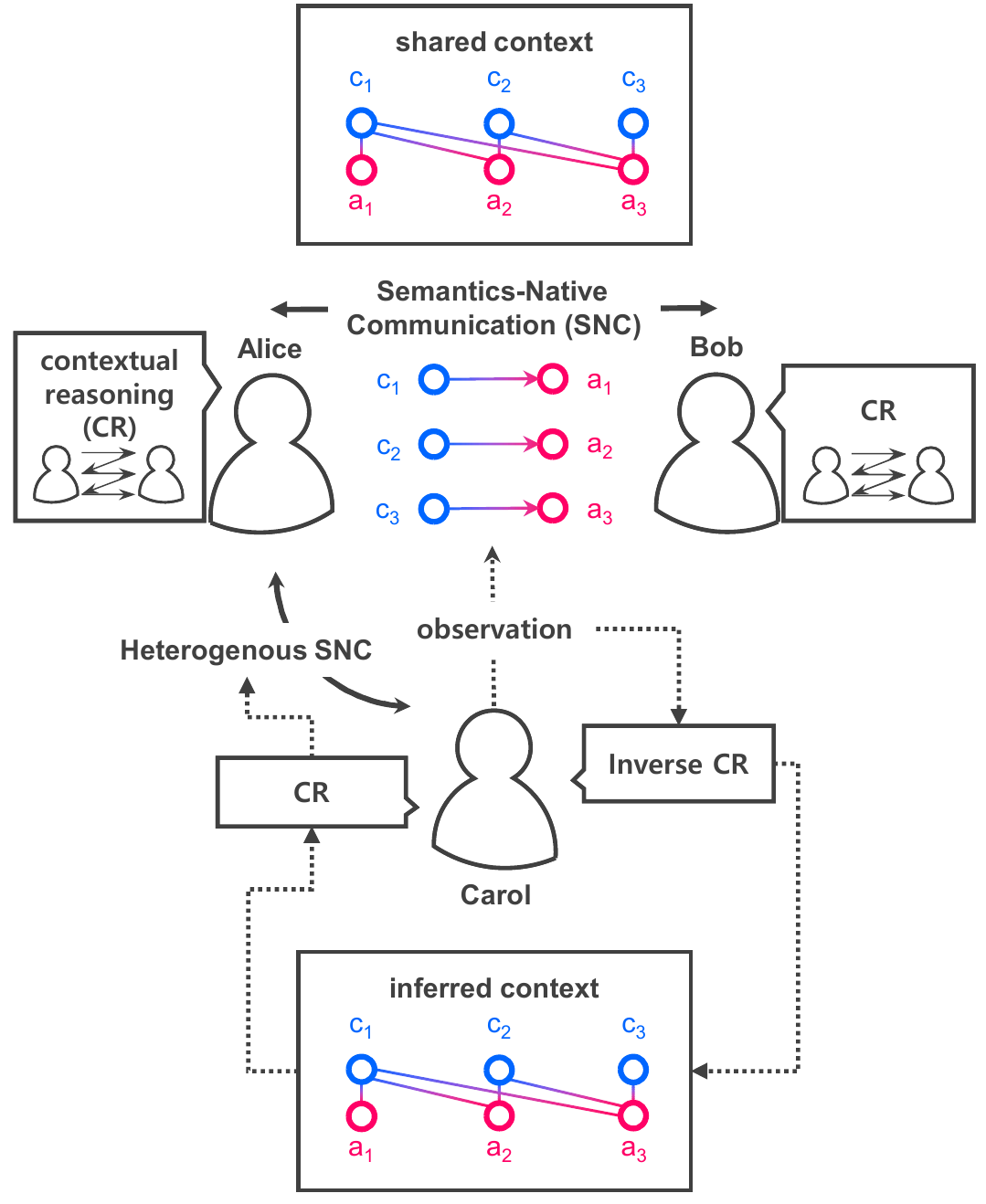}
	\caption{An illustration of the semantics-native communication (SNC) and heterogeneous SNC scenario.}
	\label{fig:system}
\end{figure}

Nonetheless, when communicating agents lack a shared context, i.e., a heterogeneous SNC scenario, CR may become less effective in achieving communication efficiency and effectiveness. To illustrate this scenario, as shown in Fig. \ref{fig:system}, consider two agents, Alice and Bob, who are involved in tasks that require coordination and share a task-oriented context characterized by mapping between task-related actions and concepts. Notions about actions and concepts will be explained in the following section. To communicate with each other and complete their tasks, Alice and Bob employ CR. Now, suppose that a third agent, Carol, is introduced into the system. Carol is unaware of Alice and Bob's communication context and the tasks they are performing. Instead, all Carol can do is observe noisy samples of communication exchanged between Alice and Bob. Therefore, Carol's goal is to understand the underlying communication context between Alice and Bob and communicate with them based on CR.\footnote{Here, Carol can also be seen as an agent who eavesdrops on Alice and Bob's communication. However, in this study, we focus not on how to prevent Carol from eavesdropping, but on how 
she can understand the context through communication samples.} To solve this challenge, Carol must first infer the context from the noisy perceived communication samples, then utilize the inferred context to communicate with Alice and Bob using CR. It is crucial to note that synchronization of context must occur before the agents can communicate effectively. Once the context is synchronized, the agents can communicate using CR even when the context changes, as long as the agents are aware of them.

This article focuses on addressing the problem of how a novel agent, Carol, can communicate effectively with other agents, Alice and Bob, who share a task-oriented context and communicate based on CR. Specifically, we propose a solution to the \emph{inverse CR (iCR)} problem, which involves extracting the ground-truth context from noisy communication samples. To tackle this problem, we propose a \emph{Bayesian iCR} method based on a \emph{two-stage Metropolis-Hastings (tMH)} algorithm, which is a Markov Chain Monte Carlo (MCMC) sampling. The tMH algorithm aims to provide samples that follow a target distribution, which in our case is the posterior distribution of inputs to the CR, i.e., the context and prior distributions of concepts and actions, given the communication samples. The algorithm is iterative, with each iteration drawing a candidate sample from a proposal distribution and accepting or rejecting it based on an acceptance probability computed from the target and proposal distributions. To overcome the difficulties of dealing with the intractable target distribution, we use a samplable standard probability distribution for proposals. The proposed algorithm is divided into two stages as the context is a sparse matrix, while the prior distributions of concepts and actions are vectors with coupled entries that sum to 1. In the first stage, the proposed scheme samples the context entries one by one to recover the sparse pattern of the matrix. Then, in the second stage, the prior distributions of concepts and actions are sampled as a whole from a probability simplex to reconstruct the vectors.

The proposed Bayesian iCR method exhibits high context inference accuracy and resultant SNC; however, its high computational complexity poses a serious challenge. The primary reason for this challenge is the computationally expensive CR required when calculating the acceptance probability in the tMH to decide whether to accept or reject a candidate sample. To overcome this challenge, we propose to linearize the CR by training a two-layer fully-connected linear neural network model. The training of the \emph{linearized CR (LCR)} model is achieved by designing a loss function that includes loss terms related to inference accuracy, SNC effectiveness, and invertibility of the CR.

On the one hand, this trained LCR model can be used instead of CR in the SNC. When LCR is used instead of the original recursive CR, it enables more efficient communication through significantly faster computations.
On the other hand, by utilizing the LCR model, the iCR problem can be transformed into an \emph{inverse LCR (iLCR)} problem, which is an under-determined linear system. As the context is a sparse matrix, the iLCR problem can also be viewed as a compressed sensing problem. To address this challenge, we propose a \emph{Bayesian iLCR} method that employs the tMH approach similar to Bayesian iCR. The LCR model significantly enhances the computational efficiency of the Bayesian iLCR method when computing the acceptance probability in tMH. Moreover, the LCR model is computationally efficient when performing SNC.

The proposed Bayesian iCR and Bayesian iLCR will be shown to effectively solve the iCR problem in heterogeneous SNC situations. In particular, we corroborate the fact that Bayesian iLCR performs better in context inference than Bayesian iCR, although pre-training is required to linearize CR. In addition, the effectiveness and efficiency of heterogeneous SNC based on the context restored using the two methods are experimentally verified.

\subsection{Related Works}
The semantic-effectiveness problem was identified a year after Shannon's mathematical communication theory was introduced \cite{Shannon1964}. Recently, with the help of advanced machine learning techniques, the problem has been re-examined through the semantic communication framework, which can be divided into three sub-directions \cite{Qin2022,Luo2022}. The first sub-direction focuses on extracting semantically important information for communication. Unlike classical compression in source coding, meaningfulness is determined not only by raw data characteristics \cite{Kalfa2021}, but also by receiver-centric metrics, such as the age-of-incorrect information (AoII) \cite{Kountouris2020,Maatouk2020}, control stability \cite{Popovski2019,Weihang2019}, and attention-based similarity \cite{Shiri2022} between agents. The second sub-direction involves transforming the raw data modality while maintaining its meaning. For instance, image-to-text embedding via Transformer \cite{Xie2021a,Xie2021b} and conversion from system states into the state evolution law using the Koopman operator \cite{Girgis2022}. Lastly, the third sub-direction involves building a knowledge base to exploit knowledge as side information or leverage dependencies in target data, which demonstrated a reduction in communication payload sizes in speech-based communication, video streaming, and holographic communication applications \cite{CalBar21,Shi2021}.

However, existing semantic communication frameworks have limitations, as they are often restricted to specific data domains (e.g., images and natural languages) or environments (e.g., control systems), and their ML-based internal operations lack a principled and explainable approach. To address these issues, \cite{Seo2023} was the first to propose a stochastic model of semantic communication based on conditional random fields, called SNC, which is compatible with the standard Shannon communication model. Nonetheless, as the SNC requires agents to share the same context to be effective and efficient, it is difficult for agents that do not know the context to participate in communication for the first time to leverage SNC. This study focuses on exploring ways to solve this challenging problem.

The concept of CR refers to the human ability to grasp implied meanings in communication, including linguistic ambiguity and communication intent, by taking into account the social interaction and local context of communication \cite{Grice1975, Bell1995, Bell1999, Bell2001}. One widely used approach for studying CR is the rational speech act (RSA) framework \cite{Frank2012, Goodman2013, Goodman2016, Frank2016}. The RSA framework models the sender and receiver as stochastic models, simulating the CR in human communication. Recently, researchers have explored the relationship between the RSA model and optimal transport in \cite{Wang2020}, and investigated it from an information-theoretic rate-distortion perspective in \cite{Zaslavsky2020}. However, these computational approaches mostly focus on the sender and overlook the receiver and their interaction with the sender. In contrast, the SNC computational framework emphasizes the CR of both the sender and receiver, as well as their interactions. While effective and efficient communication methods have been established based on CR, no known method exists to understand the context based on communication results. This article aims to address this issue by enabling agents that do not share context to perform SNC.

\subsection{Contributions and Organization}
The key contributions of this article are summarized as follows.
\begin{itemize}
    \item This work is the first of its kind that discusses and solves the inverse problem of CR-based SNC. Since SNC is known to be efficient from a communication perspective, its inverse problem is basically under-determined and difficult to solve. This article provides a couple of solutions that effectively and efficiently infer the communication context from SNC samples.
    \item To solve the iCR problem, we first introduce a Bayesian iCR method, which is based on the tMH, which is akin to MCMC sampling. The method first models CR as a likelihood function, and by using the Bayes rule, the posterior distribution can be found. Regarding the posterior distribution as a target, the tMH provides samples that follow the target distribution to enable communication with a third agent.
    \item To reduce the computational burden of the Bayesian iCR method, we linearize the CR by training a two-layer fully-connected linear neural network model. The loss function for training the model incorporates loss terms related to inference accuracy, communication effectiveness, and restricted isometry property (RIP) for CR invertibility.
    \item Based on the LCR model, we recast the iCR problem as an iLCR problem and propose a Bayesian iLCR method to solve it. The Bayesian iLCR also uses the proposed tMH but uses the trained LCR model when computing the acceptance probability, which results in lower computational complexity compared to the Bayesian iCR.
    \item We provide numerical simulation results that corroborate the communication effectiveness and computational efficiency of the proposed Bayesian iLCR. Although the Bayesian iCR is a novel approach, we regard it as a baseline and compare it with the Bayesian iLCR method. The results show that the Bayesian iLCR outperforms the Bayesian iCR method in terms of inference accuracy and computational complexity.
\end{itemize}

The rest of this article is organized as follows. In Section II, we provide preliminaries including the system model under study and background knowledge to understand SNC and MCMC sampling. In Sections III and IV, a Bayesian iCR and Bayesian iLCR are proposed, respectively. Numerical results are provided in Section V, and concluding remarks are given in Section VI.



\section{Preliminaries}\label{Sec: Preliminaries}
This section first explains the system model under study and then provides some fundamental background knowledge about semantic coding in SNC and MCMC sampling. 

\subsection{System Model}
Consider a \emph{world} wherein multiple agents deal with arbitrary tasks in the form of a referential game.\footnote{The referential game is a specific type of communication game that involves two players: a sender and a receiver. The sender is given an object, and its goal is to convey a message to the receiver so that the receiver can identify the object among a set of distractors. This game has been used to study the emergence and evolution of language, as it allows researchers to investigate how meaning is shared and how communication systems can arise from individual interactions.} To coordinate, agents communicate in pairs, each of which consists of a \emph{sender} and \emph{receiver} agents. The sender's goal is to induce the corresponding receiver to take a specific task-related action, which will hereafter be referred to as a \emph{target action}. Let $\mathbb{A}$ be the finite set of possible actions that the receiver can take, where one of them is the target and the others are distractors. We assume that there are no pre-agreed symbols representing actions among agents, and as a result, 
the agents cannot directly communicate the target action. Denote by $A\in\mathbb{A}$ the random variable of the target action chosen by the sender, and $p(A)$ its prior distribution.

In the meantime, let $\mathbb{C}$ be the finite set of \emph{concepts} that the agents commonly learn and understand while solving the given tasks. Here, a concept can be seen as a fundamental unit feature (meaning) that composes actions. Each action in $\mathbb{A}$ has one or more relevant concepts in $\mathbb{C}$. For fixed sets of actions $\mathbb{A}$ and concepts $\mathbb{C}$, \emph{context} is defined as the relevance of concepts in $\mathbb{C}$ to actions in $\mathbb{A}$. Note that the context is determined by the tasks that the agents are carrying out. Throughout this article, a context is expressed as a matrix $\mathbf{X}= (x_{c,a}) \in \mathbb{X}$, where $\mathbb{X} \subset \{0,1\}^{|\mathbb{C}|\times |\mathbb{A}|}$ is the set of all possible context configurations in the world. The entry $x_{c,a} \in [0,1]$ indicates the relevance of the concept $c$ to the action $a$, $\forall(c,a) \in\mathbb{C} \times \mathbb{A}$. For instance, $x_{c,a} = 1$ means that the concept $c$ is relevant to the action $a$ (or $c$ and $a$ are in context), while $x_{c,a} = 0$ means the concept $c$ is irrelevant to the action $a$ (or $c$ and $a$ are out of context). In general, since each action will not be associated with all concepts, the context matrix referred to here is a sparse matrix.

Suppose a sender intends to induce the corresponding receiver to take a target action $A = a^* \in \mathbb{A}$ under a communication-limited environment, in which only a symbol of one concept can be sent at a time. We suppose that there is a unique symbol representing each concept. In order to convey the communication intent (or semantics), the sender conceptualizes $a^*$ into multiple relevant concepts, chooses the most relevant one, and symbolizes it for communication purposes. Denote by $C$ the random variable of the chosen concept, and $p(C)$ its prior distribution. The symbolized concept is sent to the receiver, and upon receiving the symbol, the receiver first takes the inverse process of symbolization to retrieve the concept. Then, the receiver takes an action $\hat{a}\in\mathbb{A}$ which is the most relevant to the obtained concept. Here, what the sender and receiver technically communicate are the symbols of concepts (symbolized concepts), where the symbols can be seen as codewords in traditional communications. Indeed, the process of obtaining a symbolized concept from the intended action at the sender and its inverse process at the receiver can be jointly designed end-to-end. However, the conceptualization process is separately considered in this work to focus on the impact of CR which will be introduced later.
If the receiver's action after communication coincides with the sender's intention, i.e., $a^* = \hat{a}$, then communication is said to be effective and the meaning was conveyed correctly.\footnote{Strictly speaking, the effectiveness of communication is task-oriented, and thus, it should be designed based on the task goal that the agents are solving. Here, we assume that the communication is effective when the receiver's action $\hat{a}$ is exactly the same as the target action $a^*$.}

\subsection{Background: Semantic Coding in SNC}
The process of conceptualization is referred to as \emph{semantic encoding}, while its inverse process is referred to as \emph{semantic decoding}. Next, we explain two different semantic coding methods which are stochastic in nature. One is na\"ive semantic coding and the other is rational semantic coding. A key difference is that the latter leverages CR for more efficient and effective communication. Specifically, we focus on the CR of rational semantic coding in the rest of this article.
\subsubsection{Na\"ive Semantic Coding}
A \emph{na\"ive} pair of agents obtains and leverages a na\"ive semantic encoder and decoder. Here, the term na\"ive is used to describe the opposite notion of rational agents who leverage CR, which will be explained after. A na\"ive sender's semantic encoder is defined by a matrix $\mathbf{S}^{\scriptscriptstyle [0]} = (s^{\scriptscriptstyle [0]}_{c,a}) \in [0,1]^{|\mathbb{C}|\times|\mathbb{A}|}$, which is derived by normalizing each column vector of the context $\mathbf{X}$ as $s^{\scriptscriptstyle [0]}_{c,a} = x_{c,a}/||\mathbf{x}_{-,a}||$ for all $a\in\mathbb{A}$, where $\mathbf{x}_{-,a} = (x_{1,a}\ x_{2,a} \cdots x_{|\mathbb{C}|,a})^{\mathsf{T}}$ is the $a$-th column vector of $\mathbf{X}$. In consequence, $\mathbf{S}^{\scriptscriptstyle [0]}$ is a stochastic matrix wherein each column summing to $1$. Similarly, a na\"ive receiver's semantic decoder is defined by a matrix $\mathbf{R}^{\scriptscriptstyle [0]} = (r^{\scriptscriptstyle[0]}_{c,a}) \in [0,1]^{|\mathbb{C}|\times|\mathbb{A}|}$, where $r^{\scriptscriptstyle[0]}_{c,a} = x_{c,a}/||\mathbf{x}_{c,-}||$ for all $c\in\mathbb{C}$; $\mathbf{x}_{c,-} = (x_{c,1}\ x_{c,2} \cdots x_{c,|\mathbb{A}|})$ is the $c$-th row vector of $\mathbf{X}$, which makes $\mathbf{R}^{\scriptscriptstyle [0]}$ also a stochastic matrix wherein each row sum to $1$.

Upon obtaining $\mathbf{S}^{\scriptscriptstyle[0]}$ and $\mathbf{R}^{\scriptscriptstyle[0]}$ respectively at the sender and receiver, the agents communicate as follows. First, the sender chooses a concept that is relevant to the target action $a^*$ based on the distribution formed by the vector $\mathbf{s}_{-,a^*}^{\scriptscriptstyle[0]} = (s_{1,a^*}^{\scriptscriptstyle[0]}\  s_{2,a^*}^{\scriptscriptstyle[0]}\ \cdots\  s_{|\mathbb{C}|,a^*}^{\scriptscriptstyle[0]})^{\mathsf{T}}$. 
Simply speaking, the sender chooses that a concept $c$ is relevant to the target action $a^*$ with probability $s_{c,a^*}^{\scriptscriptstyle[0]}$. On the other communication end, the receiver carries out semantic decoding by taking an action based on the distribution formed by the vector $\textbf{r}_{c,-}^{\scriptscriptstyle[0]} = (r_{c,1}^{\scriptscriptstyle[0]}\ r_{c,2}^{\scriptscriptstyle[0]}\ \cdots\ r_{c,|\mathbb{A}|}^{\scriptscriptstyle[0]})$. In other words, the receiver takes an action $\hat{a}$ with probability $r_{c,\hat{a}}^{\scriptscriptstyle[0]}$ for some given concept $c \in\mathbb{C}$. As a matter of fact, na\"ive semantic coding is not of concern in this article. Instead, we focus on rational semantic coding explained next.

\subsubsection{Rational Semantic Coding with Contextual Reasoning}\label{subsec:pragmatic_resoning}
As mentioned earlier, leveraging context when communicating brings rationality to the agents. One prominent approach that views such rationality through the lens of a simple computational stochastic model is \emph{rational speech act model (RSA)} \cite{Frank2012, Goodman2013, Goodman2016, Frank2016}. In brief, the RSA model involves a recursive process in which the sender and receiver reason about each other's beliefs for updating their own. This resembles human reasoning when communicating with others, e.g., \emph{I think of you thinking of me thinking of you and so on} when their beliefs are commonly grounded.

Suppose that both sender and receiver are commonly grounded by the same context $\mathbf{X}$. Then, the sender can reflect the receiver's way of taking actions when designing its semantic encoder, and the receiver can do likewise for its semantic decoder. To this end, the sender and receiver independently implement a recursion that results in obtaining a rational semantic encoder and decoder. To explain, let $\mathbf{S}^{\scriptscriptstyle[t]} = (s_{c,a}^{\scriptscriptstyle[t]}) \in [0,1]^{|\mathbb{C}|\times|\mathbb{A}|}$ and $\mathbf{R}^{\scriptscriptstyle[t]} = (r_{c,a}^{\scriptscriptstyle[t]}) \in [0,1]^{|\mathbb{C}|\times|\mathbb{A}|}$ be the matrices describing the rational semantic encoder and decoder at the recursion depth $t \in \mathbb{Z}_{+}$, respectively. The entries of the rational semantic encoder and decoder matrices respectively follow the recursion \cite{Seo2023}
\begin{align}\label{eq:RSA_recursion}
\begin{cases}[\langle]
s_{c,a}^{\scriptscriptstyle [t]} = \frac{\exp\left[\theta_s u_s\left(r^{\scriptscriptstyle [t-1]}_{c,a}\right)\right]}{\sum_{c\in\mathbb{C}}\exp\left[\theta_s u_s\left(r^{\scriptscriptstyle [t-1]}_{c,a},c\right)\right]},\\
r_{c,a}^{\scriptscriptstyle[t]} = \frac{\exp\left[\theta_r u_r\left(s^{\scriptscriptstyle [t-1]}_{c,a}\right)\right]}{\sum_{a\in\mathbb{A}}\exp\left[\theta_r u_r\left(s^{\scriptscriptstyle [t-1]}_{c,a},a\right)\right]},
\end{cases}
\forall (c,a) \in\mathbb{C}\times\mathbb{A},
\end{align}
where $\theta_s \geq 0$ and $\theta_r \geq 0$ determine the rationality degree of the sender and receiver, respectively. The function $u_s(r^{\scriptscriptstyle [t-1]}_{c,a},c)$ is the sender's utility function about the receiver's semantic decoder in the previous recursion depth, and similarly, $u_r(r^{\scriptscriptstyle [t-1]}_{c,a},c)$ is the receiver's utility function about the sender's semantic encoder in the previous recursion depth.

The recursion \eqref{eq:RSA_recursion} is initialized with the na\"ive semantic decoder $\mathbf{R}^{\scriptscriptstyle [0]}$ given that $\mathbf{S}^{\scriptscriptstyle [t]}$ and $\mathbf{R}^{\scriptscriptstyle [t]}$ are also stochastic matrices for all $t\in\mathbb{Z}_+$. The way of leveraging the obtained rational semantic encoder and decoder for communication is similar to that of leveraging the na\"ive semantic encoder and decoder. For example, having a rational semantic encoder $\mathbf{S}^{\scriptscriptstyle[t]}$ at the recursion depth $t$, for given target action $a^*$, the sender supposes that a concept $c\in\mathbb{C}$ is relevant to $a^*$ with probability $s^{\scriptscriptstyle[t]}_{c,a^*}$, while the receiver randomly deconceptualizes $\hat{a}$ with probability $r_{c,\hat{a}}^{\scriptscriptstyle[t]}$ for some given concept $c \in \mathbb{C}$.


The convergence of recursion \eqref{eq:RSA_recursion} depends on how the utility functions are designed. For example, when the utility functions are defined by $u_s(r^{\scriptscriptstyle [t-1]}_{c,a}) = \log(r^{\scriptscriptstyle[t-1]}_{c,a}p(c))$ and $u_r(s^{\scriptscriptstyle [t-1]}_{c,a}) = \log(s^{\scriptscriptstyle    [t-1]}_{c,a}p(a))$, it is known that both the semantic encoder $\mathbf{S}^{\scriptscriptstyle[t]}$ and decoder $\mathbf{R}^{\scriptscriptstyle[t]}$ converge respectively to some stochastic matrices as the recursion depth approaches infinity $t\to\infty$ \cite{Zaslavsky2020}.\footnote{Especially, if the cardinalities of $\mathbb{C}$ and $\mathbb{A}$ are the same, the recursion implements the Sinkhorn-Knopp's (SK) matrix scaling algorithm \cite{Cuturi2013, Knopp1967, Sinkhorn1964} and as long as the context $\mathbf{X}$ contains sufficient non-zero entries, the encoder and decoder converge to a doubly stochastic matrix such that $\mathbf{S}^{*} = \mathbf{R}^{*}$, which is a diagonal scaling of $\mathbf{X}$.} In this article, consider such a utility function design to let the recursion \eqref{eq:RSA_recursion} converge and provide a converged rational semantic encoder and decoder $\mathbf{S}^{*}$ and $\mathbf{R}^{*}$, respectively.
Meanwhile, for given context $\mathbf{X}$ and priors $p(A)$ and $p(C)$, since the recursion \eqref{eq:RSA_recursion} is a deterministic algorithm which gives a unique solution, define by 
\begin{align}
\mathcal{S},\mathcal{R}: \mathbb{X} \times \mathbb{P}_{|\mathbb{A}|-1}\times\mathbb{P}_{|\mathbb{C}|-1}  \to [0,1]^{|\mathbb{C}|\times|\mathbb{A}|}
\end{align}
the mapping functions such that $\mathbf{S}^{*} = \mathcal{S}(\mathbf{X},p(A),p(C))$ and $\mathbf{R}^{*} = \mathcal{R}(\mathbf{X},p(A),p(C))$,
respectively, where $\mathbb{P}_M$ denotes the standard $M$-dimensional probability simplex, i.e., $\sum_{m=1}^{M+1}p_m = 1$ and $p_m \in [0,1]$ for all $m\in \{1,2,\dots,M +1\}$.

\begin{figure*}[!t]
	\centering
	\includegraphics[width=\textwidth]{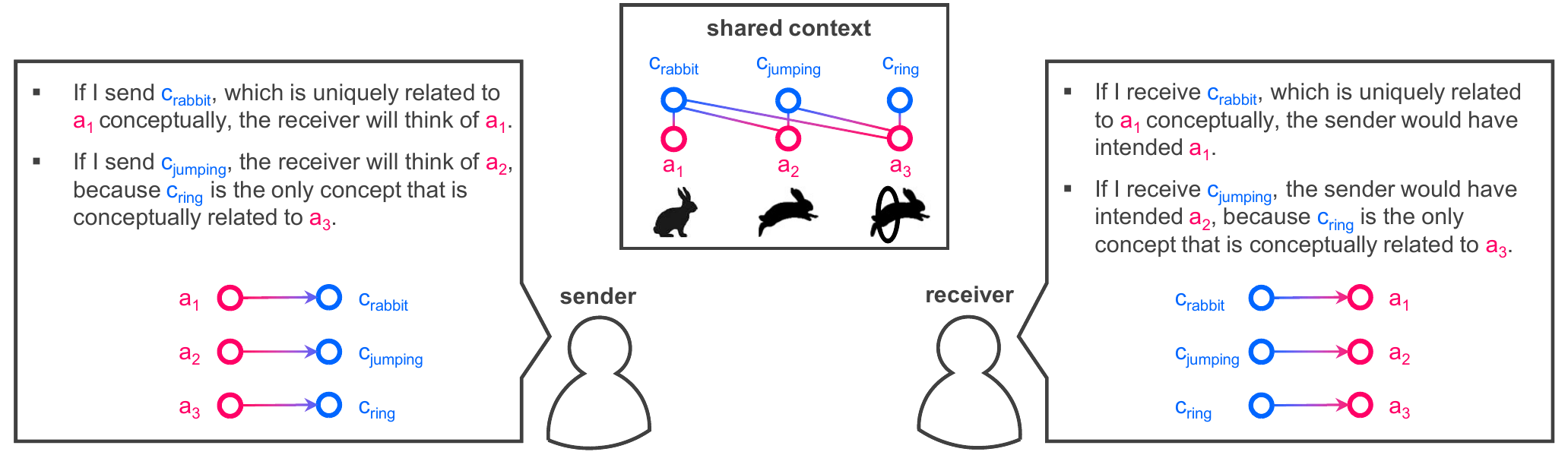}
	\caption{A referential game example with three types of actions having different concepts.}
	\label{fig:example}
\end{figure*}

\subsubsection{A Referential Game Example}
To illustrate the importance of CR, we provide the following example. Consider an instance of a world with a sender and receiver. Suppose there are three different actions that the receiver can take: pointing out to an image of a rabbit sitting ($a_1$), rabbit jumping ($a_2$), and rabbit jumping into a ring ($a_3$). In addition, there exists a set $\mathbb{C} = \{c_\text{rabbit}$, $c_\text{jumping},c_\text{ring}\}$ of the three atomic concepts that are equivalently developed at each agent as illustrated in Fig.~\ref{fig:example}. Consider a communication-limited environment in which the sender can communicate one symbolized concept at a time. Suppose the goal of the sender is to induce the receiver to take $a_2$. Under such a communication-limited environment, a na\"ive sender who should communicate two concepts relevant to $a_2$ cannot induce the receiver to take $a_2$.

By contrast, a rational sender, assuming its receiver is also rational, would select the concept $c_\text{jumping}$ to induce $a_2$. The rationale behind the selection is as follows. If the sender communicates $c_{\text{rabbit}}$, then a rational receiver will take $a_1$, since $c_{\text{rabbit}}$ is the only relevant concept of $a_1$, and thus its symbol is the most efficient and effective representation of $a_1$. Likewise, if the sender communicates the symbol of $c_{\text{ring}}$, the receiver will take $a_3$, since $c_{\text{ring}}$ is the unique concept of $a_3$. These two counter-examples justify the choice $c_\text{jumping}$ of the rational sender. Meanwhile, upon receiving the symbol of $c_{\text{jumping}}$, a na\"ive receiver cannot identify which action the sender refers to unless it additionally receives the symbol of $c_{\text{rabbit}}$. A rational receiver, by contrast, can directly take $a_2$ by reasoning in the same way as the rational sender. In conclusion, the rational agents can exchange only a single symbol of $c_{\text{jumping}}$ when referring to $a_2$, which significantly improves communication efficiency without compromising accuracy.

\subsection{Background: MCMC Sampling}
The Bayesian inference methods proposed in this article are based on MCMC sampling. Here we briefly introduce basic concepts about MCMC sampling.

\subsubsection{Monte Carlo sampling}
Monte Carlo (MC) sampling is a technique for approximating a desired quantity by extracting samples from a probability distribution. MC sampling is applicable to any probability distribution but is especially useful for intractable probability distributions in the context of inference. Representative examples of MC sampling are rejection sampling and importance sampling. Both sampling techniques generate a sample $x'$ from a simpler distribution $q(x)$ when it is difficult to obtain samples directly from a target probability distribution $p(x)$. Then, the weight assigned to the sample $x'$ obtained from the simpler distribution is set as the ratio of $q(x')$ and $p(x')$, i.e., $p(x')/q(x')$, to adjust the distribution of samples to the target distribution. For instance, by using a simpler distribution, the estimation of the expected value of a function $f(x)$ with respect to an intractable probability distribution $p(x)$ is expressed as
\begin{align}
	\mathbb{E}[f] &= \int f(x) p(x) \mathrm{d}x\\
 &= \int \frac{p(x)}{q(x)} f(x) q(x) \mathrm{d}x.
\end{align}
However, both sampling techniques rely heavily on $q(x)$ used in the sampling process, which can lead to slow convergence or inaccurate sample extraction. Moreover, these techniques are mainly applicable to univariate distributions because they require many iterations to converge.

\subsubsection{Markov Chain Monte Carlo sampling}

Markov Chain Monte Carlo (MCMC) sampling is a variant of Monte Carlo sampling that leverages Markov processes to asymptotically converge to a unique stationary distribution $\pi(x)$ that matches the target distribution $p(x)$. A Markov process is defined by its transition probability $T(x' | x)$, representing the likelihood of transitioning from one state $x$ to another state $x'$. The process achieves a unique stationary distribution $\pi(x)$ under two conditions. First, the reversibility condition (also known as detailed balance) requires that each transition $x \to x'$ is reversible, meaning $\pi(x)T(x' | x) = \pi(x')T(x | x')$. Second, the ergodicity condition ensures that the process is aperiodic (does not repeatedly return to the same state at fixed intervals) and positive recurrent (the expected number of steps to return to the same state is finite).

The Metropolis-Hastings (MH) algorithm \cite{Hastings1970,Chib1995}, a popular MCMC sampling method, involves designing a Markov process with transition probabilities that satisfy the above conditions, ultimately leading to a stationary distribution $\pi(x)$ equal to the desired target distribution $p(x)$. The algorithm's derivation begins with the reversibility condition:
\begin{align}\label{eq:reversibility}
p(x)T(x' | x) = p(x')T(x | x').
\end{align}
Similar to other Monte Carlo sampling techniques, a proposal distribution $q(x' | x)$ is introduced, representing the conditional probability of proposing state $x'$ given the current state $x$. If $q(x' | x)$ itself satisfies the reversibility condition \eqref{eq:reversibility} for all $x$ and $x'$ and symmetric $q(x'|x) = q(x|x')$, it can directly be used as the transition probability distribution ($T(x' | x) = q(x' | x)$). However, in most cases, $q(x' | x)$ does not fulfill this condition, leading to an imbalance expressed by the inequality:
\begin{align}\label{eq:unbalance}
p(x)q(x' | x) > p(x')q(x | x').
\end{align}
Informally, this indicates that the process transitions from $x$ to $x'$ too frequently, while transitioning from $x'$ to $x$ occurs too rarely. To rectify this, an acceptance probability $\alpha_{x \to x'} < 1$ is introduced to control the number of transitions from $x$ to $x'$. It should be noted that $\alpha_{x \to x'}$ is not a conditional distribution but a probability function of $x$ and $x'$. In the context of the MH algorithm, $\alpha_{x \to x'}$ is referred to as the acceptance probability in an acceptance-rejection process. Consequently, the transition probability becomes:
\begin{align}
T(x' | x) = q(x' | x) \alpha_{x \to x'}, \quad x \neq x'.
\end{align}
Returning to the inequality \eqref{eq:unbalance}, it indicates that transitions from $x'$ to $x$ are insufficient. To address this, the acceptance probability $\alpha_{x' \to x}$ is set to its maximum value, $\alpha_{x' \to x} = 1$. Applying the reversibility condition, we have:
\begin{align}
p(x)q(x' | x)\alpha_{x \to x'} &= p(x')q(x | x')\alpha_{x' \to x} \\
&= p(x')q(x | x').
\end{align}
Consequently, the MH algorithm sets the acceptance probability as:
\begin{align}
\alpha_{x \to x'} = \min\left\{1, \frac{p(x')q(x | x')}{p(x)q(x' | x)} \right\},
\end{align}
assuming $p(x)q(x' | x) \neq 0$. If the inequality in \eqref{eq:unbalance} is reversed, $\alpha_{x \to x'}$ is set to 1, and $\alpha_{x' \to x}$ is derived accordingly. It is worth mentioning that, later on, we use a notation $\alpha_{x}$ instead of $\alpha_{x \to x'}$ for the sake of brevity.

In the MH algorithm, repeated proposal and acceptance-rejection steps generate a Markov chain $(x_0, x_1, \dots, x_t, \dots)$. After a sufficient burn-in period (typically denoted by $k$ steps), the chain approaches its stationary (target) distribution $p(x)$, yielding samples $(x_{k+1}, x_{k+2}, \dots) \sim p(x)$. Compared to other MC sampling methods, the MH algorithm and MCMC sampling, in general, offer the advantage of approximating intractable and multivariate probability distributions.

\section{Bayesian Inverse Contextual Reasoning}\label{sec:BiCR}
Now, we address the problematic scenario presented earlier in Section \ref{sec:introduction}. We first restate the problem in terms of the rational semantic coding described in Section \ref{subsec:pragmatic_resoning}. Suppose Alice and Bob are grounded in the same context $\mathbf{X}$ and communicate based on the rational semantic encoder and decoder. Carol is not aware of $\mathbf{X}$, but can only collect samples from SNC between Alice and Bob. Here, one sample consists of a symbolized concept of Alice communicating with Bob and the corresponding action taken by Bob upon receiving it. Then, the problem is reduced to finding a method for Carol to estimate the context $\hat{\mathbf{X}}$, and priors $\hat{p}(A)$ and $\hat{p}(C)$ which well-approximate the ground truth $\mathbf{X}$, $p(A)$ and $p(C)$, respectively. To address such a problem, we propose a method dubbed \emph{Bayesian iCR} that infers the context and prior distributions from SNC samples based on Bayesian inference and MCMC sampling. To this end, we assume that $\mathbf{X}$, $p(A)$ and $p(C)$ are independent random variables throughout this section, and use vector notations $\boldsymbol{y} = p(A)$ and $\boldsymbol{z} = p(C)$ to avoid confusion and for simplicity. In addition, we let $\mathbf{T} = (\mathbf{X},\boldsymbol{y},\boldsymbol{z})$ be the tuple of random variables $\mathbf{X}$, $\boldsymbol{y}$ and $\boldsymbol{z}$, for the sake of convenience.

To begin with, suppose that Carol has obtained a set of noisy data samples $\mathbb{D} = \{\tilde{c}_d,\tilde{a}_d\}_{d=1}^{D}$ from observing what Alice sends and how Bob reacts. Here, `noisy' encompasses all errors that may occur during the communication process between Alice and Bob or during Carol's observation process. By counting the number of each concept-action pair in the samples set, and dividing it by the total number of samples $D$, Carol obtains an empirical decoder $\bar{\mathbf{R}}$, which is corrupted by a zero-mean additive Gaussian random noise matrix $\mathbf{N} \in \mathbb{R}^{|\mathbb{C}|\times|\mathbb{A}|}$ with i.i.d. entries of variance $\sigma^{2}$, that is
\begin{align}\label{eq:noisy_estimate}
\bar{\mathbf{R}} = \underbrace{\mathcal{R}(\mathbf{T})}_{\ \mathbf{R}^*} + \mathbf{N}.
\end{align}
Such an assumption depicts that $\bar{\mathbf{R}}$ is structured based on a large number of samples thereby following the Central Limit Theorem (CLT). The likelihood of the empirical decoder $\bar{\mathbf{R}}$ is
\begin{align}
    p(\bar{\mathbf{R}}|\mathbf{T}) &= p(\bar{\mathbf{R}}|\mathcal{R}(\mathbf{T}),\mathbf{T})\label{eq:deterministicalgo}\\
    &= p(\mathbf{N}|\mathcal{R}(\mathbf{T}),\mathbf{T})\label{eq:additiveN}\\
    &= p(\mathbf{N})\label{eq:independentnoise}\\
    &\propto \exp\left( -\frac{1}{2\sigma^2} \|\textbf{vec}(\mathbf{N})\|_2^2 \right)\label{eq:matrixnormal}
\end{align}
where \eqref{eq:deterministicalgo} holds since $\mathbf{R}^* = \mathcal{R}(\mathbf{T})$ is deterministic when $\mathbf{T}$ is given; \eqref{eq:additiveN} holds from \eqref{eq:noisy_estimate}; \eqref{eq:independentnoise} holds since the noise is independently generated; and \eqref{eq:matrixnormal} is from the probability density function of multivariate normal distribution while $\textbf{vec}(\mathbf{N})$ is the vectorization of $\mathbf{N}$. Moreover, from Bayes' rule and \eqref{eq:independentnoise}, the posterior distribution of $\mathbf{T}$ is
\begin{align}\label{eq:posterior}
    p(\mathbf{T}|\bar{\mathbf{R}}) &= \frac{p(\mathbf{N})p(\mathbf{T})}{p(\bar{\mathbf{R}})}\\
     &= \frac{1}{U}\exp\left( -\frac{1}{2\sigma^2} \|\textbf{vec}(\mathbf{N})\|_2^2 \right)p(\mathbf{T})\\
     &= \frac{1}{U}\exp\left( -\frac{1}{2\sigma^2} \textbf{vec}(\bar{\mathbf{R}}- \mathcal{R}(\mathbf{T})  )\|_2^2 \right)p(\mathbf{T}),\label{eq:posterior_2}
\end{align}
where $U = \int \exp\left( -\frac{1}{2\sigma^2} \|\textbf{vec}(  \bar{\mathbf{R}}- \mathcal{R}(\mathbf{T})  )\|_2^2 \right)p(\mathbf{T}) d\mathbf{T}$ is a normalization term and \eqref{eq:posterior_2} holds from \eqref{eq:noisy_estimate}.

Having the posterior distribution \eqref{eq:posterior_2}, one promising approach for iCR is the maximum a posteriori (MAP) point estimation. In detail, given that the prior $p(\mathbf{T})$ is known and tractable, the MAP method finds the point which maximizes the posterior, i.e.,
\begin{align}\label{eq:MAP}
    \hat{\mathbf{T}}_{\mathrm{MAP}} = \underset{\mathbf{T}}{\arg \max}\  p(\mathbf{T}|\bar{\mathbf{R}}).
\end{align}
However, in general, it is difficult to obtain the MAP estimate if the prior $p(\mathbf{T})$ is unknown or intractable.
As an alternative approach for Bayesian iCR, sampling from the numerator of \eqref{eq:posterior_2} can approximate the posterior distribution and also provide a point estimate. Specifically, we propose a two-stage Metropolis-Hastings (tMH) algorithm as shown in Algorithm \ref{alg:tMH}. In brief, the algorithm finds $\mathbf{X}$ in the first stage by applying the MH method entry-by-entry and then finds $\boldsymbol{y}$ and $\boldsymbol{z}$ in the second stage as a whole. The overall target distribution of the algorithm is the posterior distribution \eqref{eq:posterior_2} and we use an overall proposal distribution $q(\mathbf{T}'|\mathbf{T})$, where $\mathbf{T}' = (\mathbf{X}',\boldsymbol{y}',\boldsymbol{z}')$ is a random variable denoting an update proposal of $\mathbf{T}$. The proposal distribution is chosen among a list of standard samplable probability distributions. Typically, in the MH algorithm, an index representing the iteration number is used, but for the sake of simplicity and clarity in notation, this article will omit it. The sampling procedures in each stage are detailed as follows.

\subsubsection{Stage 1 (Sparse Pattern Reconstruction of $\mathbf{X}$)} 
In the first stage of the algorithm, each entry of $\mathbf{X}$ is updated one by one based on the premise that the other entries are true values. To this end, when updating $x_{c,a}$, other entries in $\mathbf{T}$ are fixed so that the target posterior distribution reduces to $p(x_{c,a}|\bar{\mathbf{R}},\mathbf{T}{\backslash x_{c,a}})$, $\forall (c,a) \in \mathbb{C}\times\mathbb{A}$,
where $\mathbf{T}{\backslash x_{c,a}}$ denotes the entries of $\mathbf{T}$ except for $x_{c,a}$. Similarly, the proposal distribution becomes $q(x'_{c,a}|\mathbf{T})$, $\forall (c,a) \in \mathbb{C}\times\mathbb{A}$, where $x'_{c,a}$ is the update proposal of $x_{c,a}$.

The procedure of updating $x_{c,a}$ can be briefly described as follows. First, $x_{c,a}'$ which follows $q(x_{c,a}'|\mathbf{T})$ is generated. Then, $x_{c,a}'$ is accepted as an update of $x_{c,a}$ with probability
\begin{align}
    \alpha_{x_{c,a}} = \min\left(1, \frac{p(x_{c,a}'|\bar{\mathbf{R}},\mathbf{T}\backslash x_{c,a})}{p(x_{c,a}|\bar{\mathbf{R}},\mathbf{T}\backslash x_{c,a})}
    \frac{q(x_{c,a}|\mathbf{T}\backslash x_{c,a}, x_{c,a}')}{q(x_{c,a}'|\mathbf{T})} \right),
\end{align}
and rejected otherwise. Each entry of $\mathbf{X}$ is updated for $K_1 \geq 1$ iterations in the first stage of the algorithm. The notation $\alpha_{x_{c,a}}$ is used instead of $\alpha_{x_{c,a}\to x_{c,a}'}$ for the sake of brevity.

\subsubsection{Stage 2 (Reconstruction of $\boldsymbol{y}$ and $\boldsymbol{z}$)}
In the second stage of the algorithm $\boldsymbol{y}$ and $\boldsymbol{z}$ are updated. In contrast to the first stage, both $\boldsymbol{y}$ and $\boldsymbol{z}$ are updated as a whole. Thus, the target distributions for $\boldsymbol{y}$ and $\boldsymbol{z}$ are $p(\boldsymbol{y}|\bar{\mathbf{R}},\mathbf{X},\boldsymbol{z})$ and $p(\boldsymbol{z}|\bar{\mathbf{R}},\mathbf{X},\boldsymbol{y})$, respectively. Similarly, the proposal distributions for updating $\boldsymbol{y}$ and $\boldsymbol{z}$ are $q(\boldsymbol{y}'|\mathbf{T})$ and $q(\boldsymbol{z}'|\mathbf{T})$, respectively.

Update candidates $\boldsymbol{y}'$ and $\boldsymbol{z}'$ are generated by sampling from the proposal distributions $q(\boldsymbol{y}'|\mathbf{T})$ and $q(\boldsymbol{z}'|\mathbf{T})$, respectively, and their acceptance probabilities are
\begin{align}
\alpha_{\boldsymbol{y}} = \min\left(1, \frac{p(\boldsymbol{y}'|\bar{\mathbf{R}},\mathbf{X},\boldsymbol{z})}{p(\boldsymbol{y}|\bar{\mathbf{R}},\mathbf{X},\boldsymbol{z})}\frac{q(\boldsymbol{y}|\mathbf{X},\boldsymbol{y}',\boldsymbol{z})}{q(\boldsymbol{y}'|\mathbf{T})} \right),
\end{align}
and
\begin{align}
\alpha_{\boldsymbol{z}} = \min\left(1, \frac{p(\boldsymbol{z}'|\bar{\mathbf{R}},\mathbf{X},\boldsymbol{y})}{p(\boldsymbol{z}|\bar{\mathbf{R}},\mathbf{X},\boldsymbol{y})}\frac{q(\boldsymbol{z}|\mathbf{X},\boldsymbol{y},\boldsymbol{z}')}{q(\boldsymbol{z}'|\mathbf{T})} \right),
\end{align}
respectively. Both $\boldsymbol{y}$ and $\boldsymbol{z}$ are updated for $K_2 \geq 1$ iterations in the second stage of the algorithm.

\begin{algorithm*}[t]
\caption{Two-Stage Metropolis-Hastings (tMH) for Bayesian iCR and iLCR}\label{alg:tMH}
\text{Set target and proposal distribution:} $p(\mathbf{T}|\bar{\mathbf{R}})$ and $q(\mathbf{T}'|\mathbf{T})$ respectively.\\
\text{Initialize:} $\mathbf{T} = (\mathbf{X}, \boldsymbol{y}, \boldsymbol{z})$
\begin{algorithmic}
\For{$K$ iterations}
\For{$K_1$ iterations}\Comment{\emph{Stage 1. Sparse Pattern Recovery of $\boldsymbol{x}$}}
\For{$1 \leq c \leq |\mathbb{C}|$ and $1 \leq a \leq |\mathbb{A}|$}
\State \textbf{Generate} $x_{c,a}' \sim q(x_{c,a}'|\mathbf{T})$
\State \quad \quad \quad \quad \  $x_{c,a} = \begin{cases}
x_{c,a}'\ \textit{(accept)} & \text{with probability } \alpha_{x_{c,a}}\\
x_{c,a}\  \textit{(reject)} & \text{otherwise}
\end{cases}$
\EndFor
\EndFor
\For{$K_2$ iterations}\Comment{\emph{Stage 2. Recovery of $\boldsymbol{y}$ and $\boldsymbol{z}$}}
\State \textbf{Generate} $\boldsymbol{y}' \sim q(\boldsymbol{y}'|\mathbf{T})$
\State \quad \quad \quad \quad \  $\boldsymbol{y} = \begin{cases}
\boldsymbol{y}'\ \textit{(accept)} & \text{with probability } \alpha_{\boldsymbol{y}}\\
\boldsymbol{y}\  \textit{(reject)} & \text{otherwise}
\end{cases}$
\State \textbf{Generate} $\boldsymbol{z}' \sim q(\boldsymbol{z}'|\mathbf{T})$
\State \quad \quad \quad \quad \  $\boldsymbol{z} = \begin{cases}
\boldsymbol{z}'\ \textit{(accept)} & \text{with probability } \alpha_{\boldsymbol{z}}\\
\boldsymbol{z}\  \textit{(reject)} & \text{otherwise}
\end{cases}$
\EndFor
\EndFor
\end{algorithmic}
\end{algorithm*}

The first and second stages are repeated $K\geq 1$ times. As mentioned, one of the advantages of the proposed algorithm is that it works regardless of whether the posterior distribution is tractable or not. It is worth noting that when we have a samplable posterior distribution, we can apply the Gibbs sampling method \cite{Geman1984} instead. The algorithm has been divided into two stages due to the nature of the input data. Specifically, the context is represented as a sparse matrix, while the prior distributions of the concept and action are represented as vectors, in which the entries are coupled to each other and the sum of the entries is constrained to be unity. By dividing the algorithm into two stages, it becomes possible to handle these different types of data in a more efficient and effective manner. Moreover, it enables the design of different proposal distributions for different components and overcomes the curse of dimensionality.

\section{Bayesian Inverse Linearized Contextual Reasoning}\label{sec:BiLCR}
One critical problem with the SNC-based framework using CR is that running the iterative recursion \eqref{eq:RSA_recursion} until its convergence is computationally expensive. This problem becomes more pronounced in the Bayesian iCR introduced in the previous section because the CR must be repeatedly done for the accept/reject processes in tMH. One way to avoid such a problem is to linearize the CR to obtain the rational encoder and decoder without recursive and iterative computation. In this section, we propose a \emph{Bayesian iLCR} method on the basis of the LCR model obtained from training a two-layer fully-connected neural network and the tMH.

\subsection{Linearizing Contextual Reasoning}
Assuming that the mapping function $\mathcal{R}$ is linearized, the rational semantic decoder can be written as
\begin{align}\label{eq:linearize}
    \mathbf{vec}(\tilde{\mathcal{R}}(\mathbf{T})) &= \mathbf{\Phi}\mathbf{vec}(\mathbf{T})
\end{align}
for some $|\mathbb{C}||\mathbb{A}| \times (|\mathbb{C}||\mathbb{A}|+|\mathbb{C}|+|\mathbb{A}|)$ dimensional matrix $\mathbf{\Phi}$, where $\tilde{\mathcal{R}}(\mathbf{T}) = (\tilde{r}_{c,a}) \in [0,1]^{|\mathbb{C}|\times|\mathbb{A}|}$ is the matrix that approximates $\mathcal{R}(\mathbf{T})$ incorporating the impact of perturbation that comes from the linearization. In addition, $\mathbf{vec}(\tilde{\mathcal{R}}(\mathbf{T})) \in \mathbb{R}^{|\mathbb{C}||\mathbb{A}|}$ and $\mathbf{vec}(\mathbf{T}) \in \mathbb{R}^{|\mathbb{C}||\mathbb{A}|+|\mathbb{C}|+|\mathbb{A}|}$ are the column vectors obtained by vectorizing $\tilde{\mathcal{R}}(\mathbf{T})$ and $\mathbf{T}$, respectively. The rational semantic encoder obtained through a single iteration of $\eqref{eq:RSA_recursion}$ upon $\tilde{\mathcal{R}}(\mathbf{T})$ is denoted by $\tilde{\mathcal{S}}(\mathbf{T}) = (\tilde{s}_{c,a}) \in [0,1]^{|\mathbb{C}|\times|\mathbb{A}|}$.

One effective way to obtain the matrix $\mathbf{\Phi}$ is to train a two-layer fully-connected  neural network without non-linear activations which takes $\mathbf{T}$ as input and $\mathcal{R}(\mathbf{T})$ as output. In general, it is difficult to train such a matrix that linearizes a given recursive non-linear evolution with small misfit  $\|\mathbf{vec}(\mathcal{R}(\mathbf{T}) - \tilde{\mathcal{R}}(\mathbf{T}))\|_2^2$. However, in the scenario under consideration, it is possible to obtain $\mathbf{\Phi}$ since both $\mathbf{vec}(\mathcal{R}(\mathbf{T}))$ and $\mathbf{vec}(\mathbf{T})$ are sparse vectors from the structure. Later on, we show that whether the recursion is linearizable or not depends on the sparsity of the context $\mathbf{X}$. Moreover, when training $\mathbf{\Phi}$ (or designing loss function for training the linear neural network model), our focus is not only on minimizing the misfit $\|\mathbf{vec}(\mathcal{R}(\mathbf{T}) - \tilde{\mathcal{R}}(\mathbf{T}))\|_2^2$ but also ensuring the communication effectiveness of the approximated rational semantic coding pair $(\tilde{\mathcal{S}}(\mathbf{T}), \tilde{\mathcal{R}}(\mathbf{T}))$, as much as the original pair $(\mathcal{S}(\mathbf{T}),\mathcal{R}(\mathbf{T}))$ provides.

The LCR model is trained as follows. First, generate data samples $(\mathbf{T}_i,\mathcal{R}(\mathbf{T})_i)_{i=1}^{N}$ for a fixed prior distribution, i.e., $\mathbf{T}_i \sim p(\mathbf{T})$, $1\leq \forall i \leq N$, and then obtain the neural network model weights that minimize the loss function defined by
\begin{align}\label{eq:lossfunction_1}
    \mathcal{L}_{1} = \sum_{i=1}^{N} \left\{\lambda_{\text{mis}}\mathcal{L}_{\text{mis}}(\mathbf{T}_i) + \lambda_{\text{eff}} \mathcal{L}_{\text{eff}}(\mathbf{T}_i)\right\}
\end{align}
for some constants $\lambda_{\text{mis}}, \lambda_{\text{eff}} \geq 0$ such that $\lambda_{\text{mis}} + \lambda_{\text{eff}} = 1$, where $\mathcal{L}_{\text{mis}}(\mathbf{T}) = \| \mathbf{vec}(\mathcal{R}(\mathbf{T}) - \tilde{\mathcal{R}}(\mathbf{T}))\|_2^2$ and $\mathcal{L}_{\text{eff}}(\mathbf{T}) = \sum_{a\in\mathbb{A}} p(a) \left(1 - \sum_{c\in\mathbb{C}} \tilde{s}_{c,a}\tilde{r}_{c,a}\right)$ are misfit and effectiveness loss term, respectively. It is worth mentioning that the loss function can be defined in various ways, for example using cross-entropy for each row vector of $\mathcal{R}(\mathbf{T})$ and $\tilde{\mathcal{R}}(\mathbf{T})$, however, we used mean squared error (MSE) misfit function to obtain the linearized form of the CR in \eqref{eq:linearize}. The original CR model does not have a one-to-one relationship between input and output. However, the LCR model can be seen as a one-to-one mapping function. Therefore, from the perspective of invertibility, it can be understood that utilizing the LCR model instead of the CR model is more advantageous. On the other hand, in a situation where Carol acts as the eavesdropper and Alice and Bob aim to prevent eavesdropping, it is more advantageous to use CR.

\subsection{Invertible Linearized Contextual Reasoning}
Recall that the dimension of $\mathbf{vec}(\mathbf{T})$ is larger than that of $\mathbf{vec}(\tilde{\mathcal{R}}(\mathbf{T}))$, which in turn makes the linear system \eqref{eq:linearize} underdetermined. In addition, note that the entries of $\mathbf{T}$ (strictly saying $\mathbf{X}$) are sparse. In consequence, having the LCR model, the iCR problem is recast as an iLCR problem (or compressed sensing problem), where $\mathbf{vec}(\tilde{\mathcal{R}}(\mathbf{T}))$ is a noiseless observation vector, $\mathbf{\Phi}$ is a measurement matrix, and $\mathbf{vec}(\mathbf{T})$ is an unknown vector which requires to be reconstructed. Considering the additive noise in \eqref{eq:additiveN}, a noisy empirical decoder $\bar{\mathbf{R}}$ corrupted by a zero-mean additive Gaussian random noise $\mathbf{N} \in \mathbb{R}^{|\mathbb{C}|\times|\mathbb{A}|}$ is $\bar{\mathbf{R}} = \tilde{\mathcal{R}}(\mathbf{T}) + \mathbf{N}$, and in a vectorized form
\begin{align}
    \mathbf{vec}(\bar{\mathbf{R}}) &= \mathbf{vec}(\tilde{\mathcal{R}}(\mathbf{T}) + \mathbf{N})\\
    &= \mathbf{\Phi}\mathbf{vec}(\mathbf{T}) + \mathbf{vec}(\mathbf{N}).\label{eq:noisylinear}
\end{align}
For the sake of convenience, instead of notations $\mathbf{vec}(\bar{\mathbf{R}})$, $\mathbf{vec}(\mathbf{T})$, $\mathbf{vec}(\mathbf{X})$ and $\mathbf{vec}(\mathbf{N})$, we use vectors $\bar{\boldsymbol{r}}$, $\boldsymbol{t}$, $\boldsymbol{x}$ and $\boldsymbol{n}$ throughout this section, thereby having the following simplified equation
\begin{align}
    \bar{\boldsymbol{r}} = \mathbf{\Phi}\boldsymbol{t} + \boldsymbol{n},
\end{align}
which is equivalent to \eqref{eq:noisylinear}.

Now, suppose that $\boldsymbol{t}$ is a $s$-sparse vector, which means that the number of non-zero entries in the vector is (at most) $s \in \mathbb{Z}_+$. Then, according to the compressed sensing theory, the matrix $\mathbf{\Phi}$ should satisfy the $(s,\delta)$-restricted isometry property (RIP) condition in order to correctly estimate the unknown vector. The matrix $\mathbf{\Phi}$ is said to satisfy the $(s,\delta)$-RIP if there exists $\delta \in (0,\frac{1}{3})$ such that every $s$-sparse vector $\boldsymbol{v}$ holds
\begin{align}\label{eq:RIP}
(1-\delta)\|\boldsymbol{v}\|_2^2 \leq \|\mathbf{\Phi}\boldsymbol{v}\|_2^2 \leq (1+\delta)\|\boldsymbol{v}\|_2^2.
\end{align}
In normal terms, $\mathbf{\Phi}$ behaves like an isometry (orthonormal) matrix, even though it is not a square matrix. As a matter of fact, obtaining $\mathbf{\Phi}$ that meets the RIP condition \eqref{eq:RIP} is a strongly NP-hard problem \cite{Tillmann2014}. Moreover, the matrix $\mathbf{\Phi}$ in our case must also linearize the CR, which makes it harder to obtain such a matrix. However, it is known that some random matrices, for example, the Gaussian or Bernoulli distribution, satisfy the RIP condition with high probability. Considering that the linear neural network for $\mathbf{\Phi}$ that minimizes $\mathcal{L}_1$ is trained based on a stochastic optimization such as stochastic gradient descent (SGD) method under Gaussian random initialization, it is likely that the trained $\mathbf{\Phi}$ meets the $(s,\delta)$-RIP condition.

Nevertheless, in order to further ensure an invertible CR, we propose an alternative loss function that incorporates the $(s,\delta)$-RIP constraint to train the linear neural network. The proposed loss function can be written as
\begin{align}\label{eq:lossfunction_2}
    \mathcal{L}_{2} = \underbrace{\mathcal{L}_{1}}_{\eqref{eq:lossfunction_1}} + \lambda_2 \sum_{i=1}^{N}\mathcal{L}_{\text{RIP}}(\mathbf{T}_i),
\end{align}
for some constant $\lambda_2 \geq 0$, where $\mathcal{L}_{\text{RIP}} = \left(\frac{\|\mathbf{\Phi}\boldsymbol{t}\|_2^2}{\|\boldsymbol{t}\|_2^2}-1\right)^2$. When minimizing $\mathcal{L}_{2}$, the new regularization term forces the $\mathbf{\Phi}$ to satisfy \eqref{eq:RIP}, for all vectorized data samples $\mathbf{T}_1,\mathbf{T}_2,\dots,\mathbf{T}_N$, which we call a quasi-RIP condition. Since it does not fully characterize the original $(s,\delta)$-RIP condition, in the sense that the condition only holds for the data samples, the regularization term still cannot fully guarantee that the matrix $\mathbf{\Phi}$ trained by using \eqref{eq:lossfunction_2} holds the $(s,\delta)$-RIP condition. However, compared to that trained with $\mathcal{L}_{1}$, it is more likely to hold the condition if the constant $\lambda_2$ is chosen appropriately.

\begin{figure*}[t!]
\centering
\subfigure[]{\includegraphics[width=0.329\textwidth]{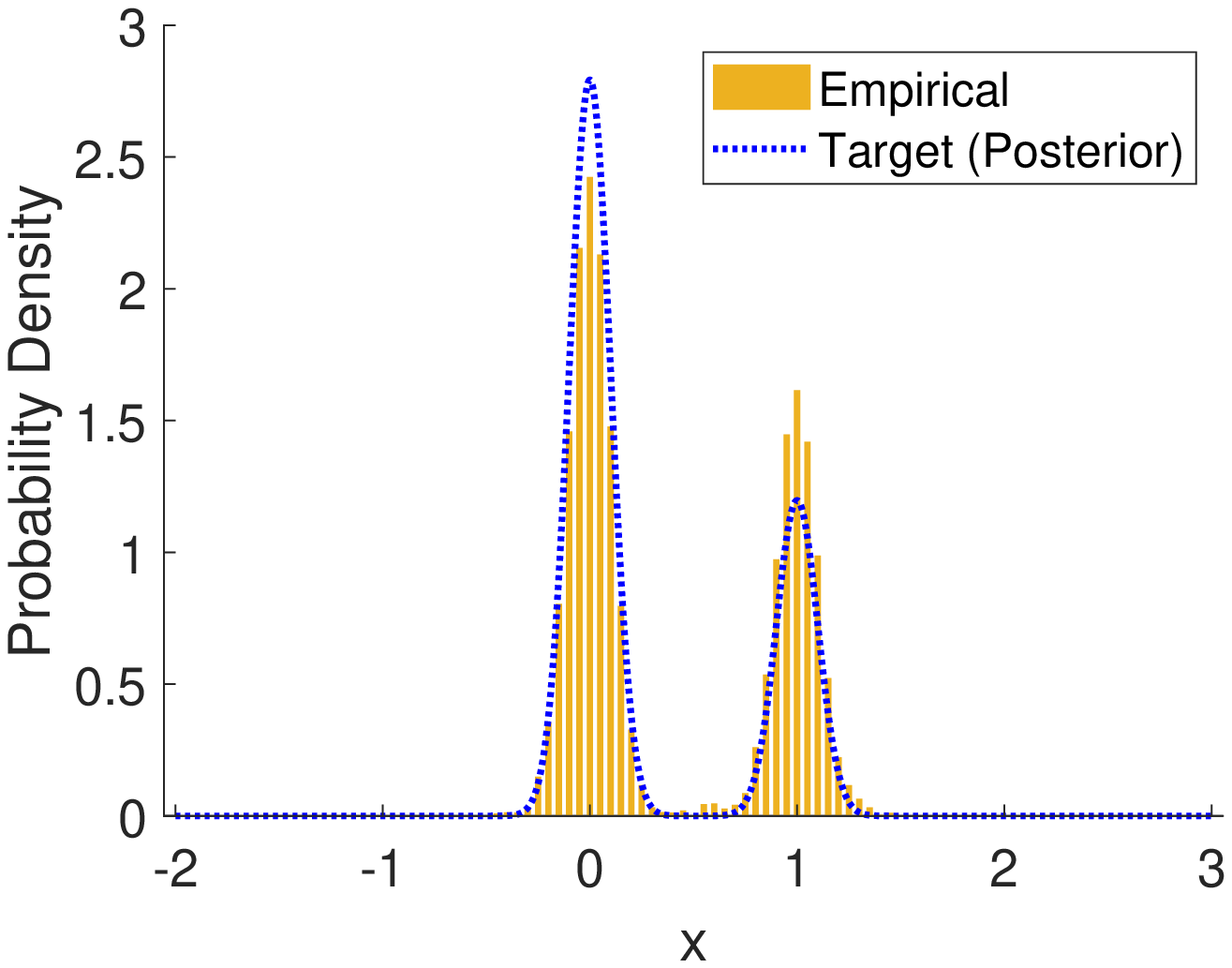}\label{fig:sim_1_1}}
\subfigure[]{\includegraphics[width=0.329\textwidth]{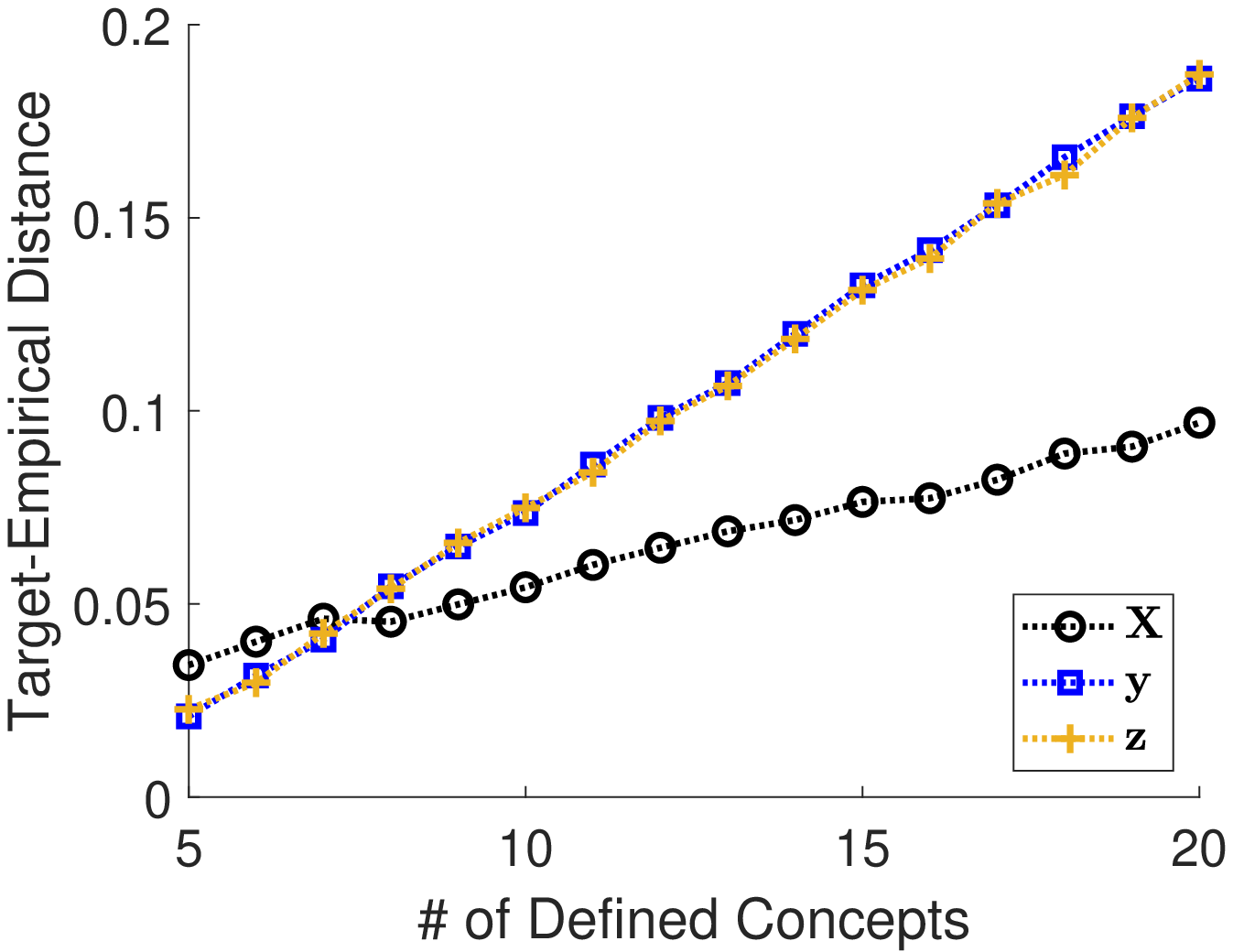}\label{fig:sim_1_2}}
\subfigure[]{\includegraphics[width=0.329\textwidth]{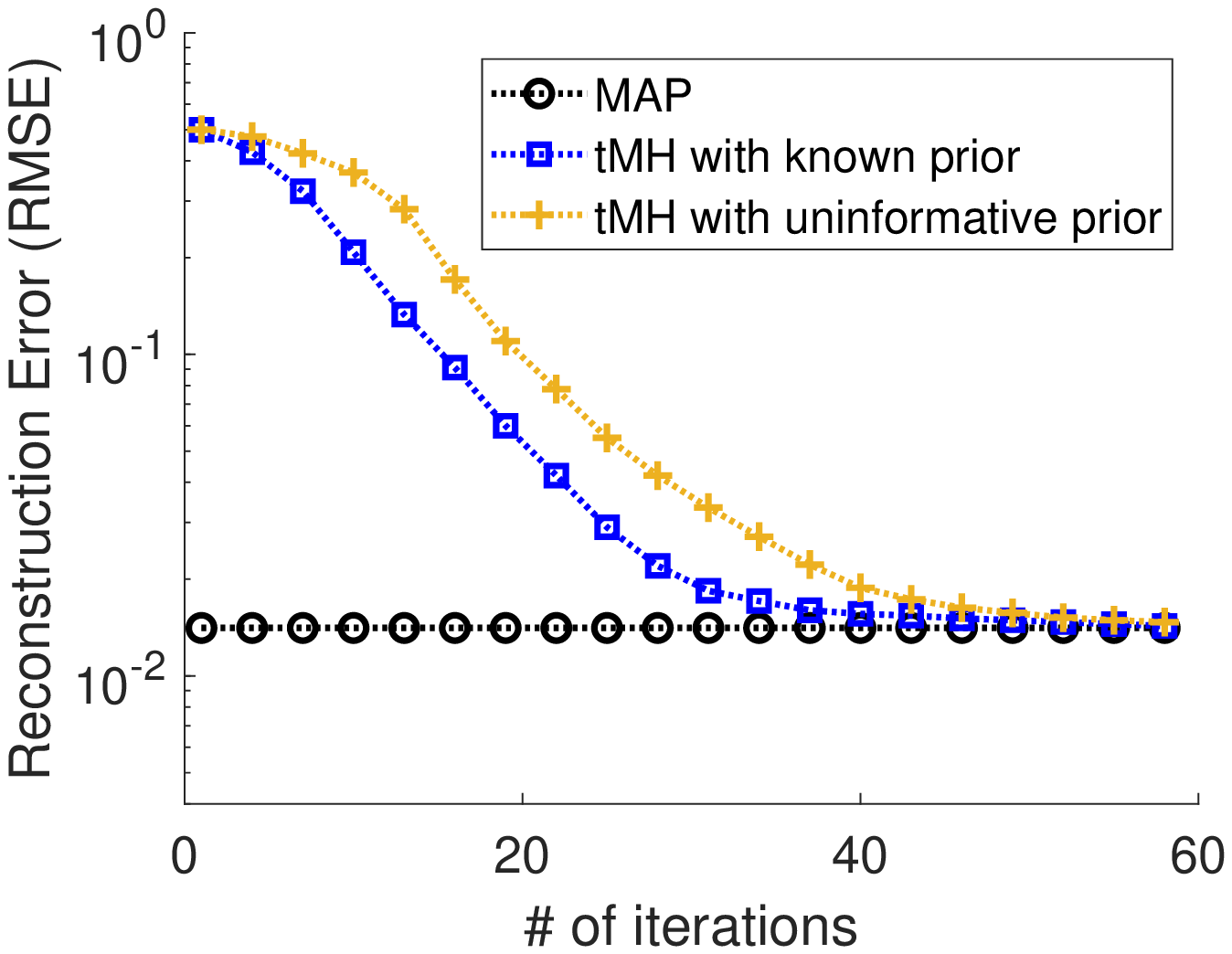}\label{fig:sim_1_3}}
\subfigure[]{\includegraphics[width=0.329\textwidth]{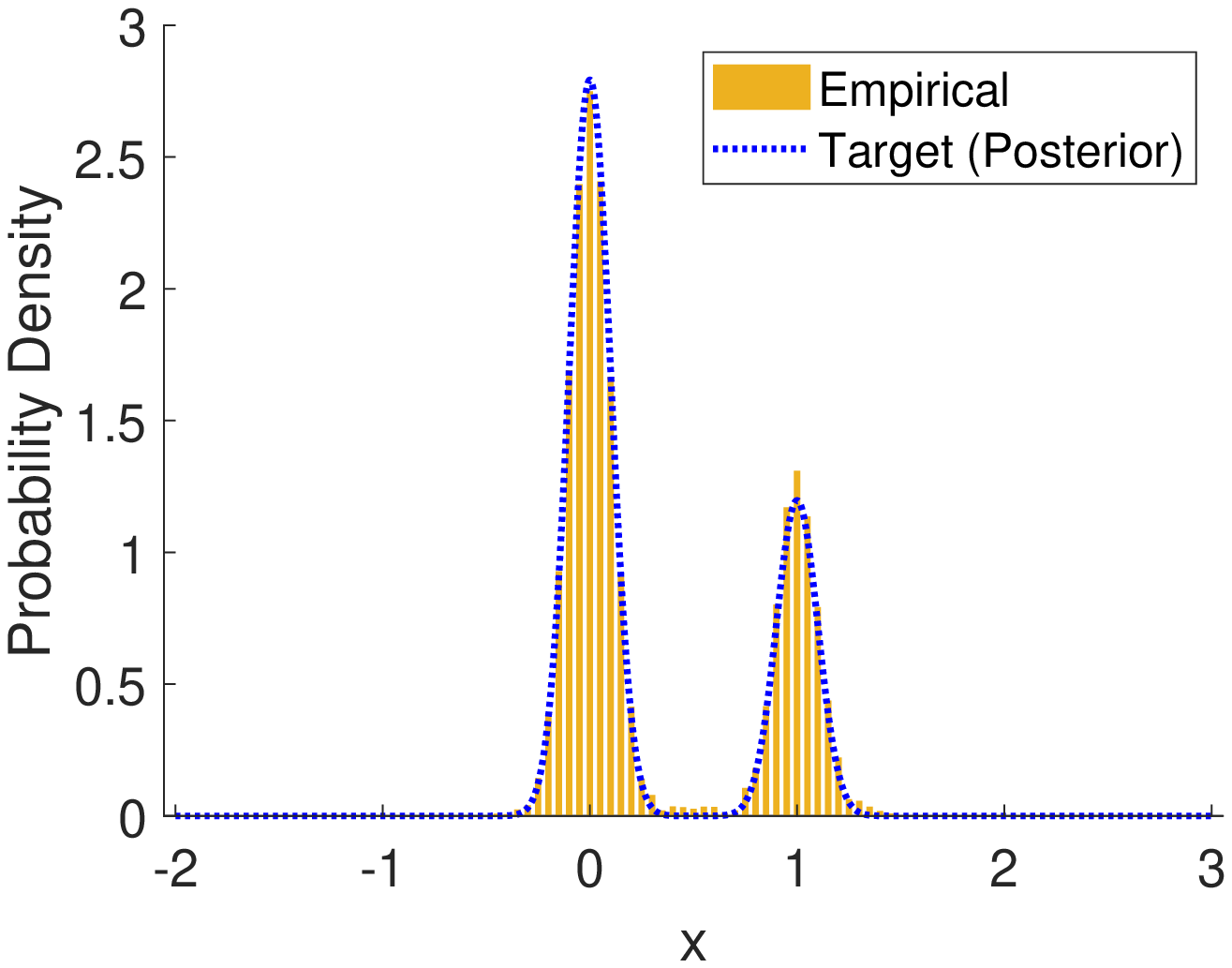}\label{fig:sim_2_1}}
\subfigure[]{\includegraphics[width=0.329\textwidth]{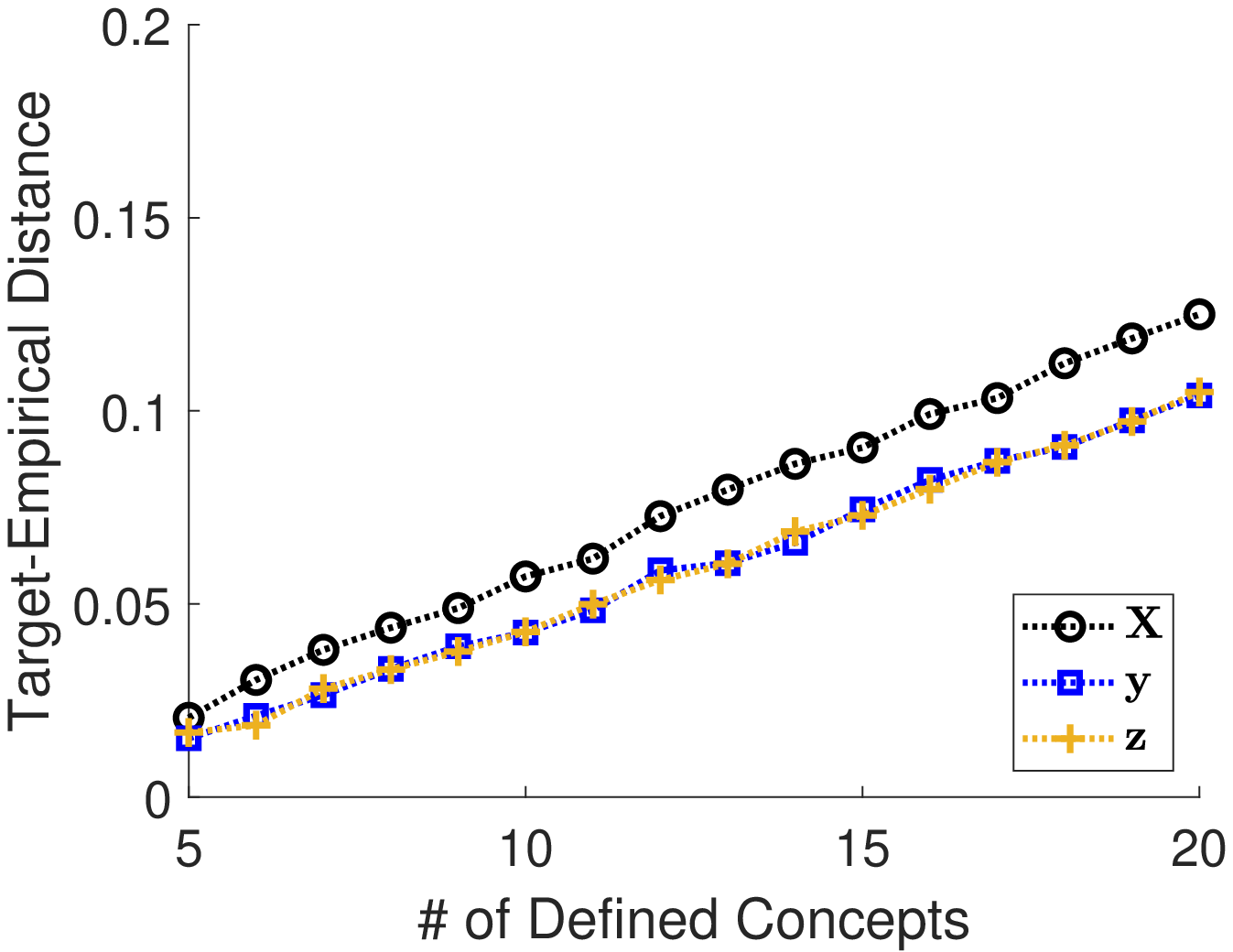}\label{fig:sim_2_2}}
\subfigure[]{\includegraphics[width=0.329\textwidth]{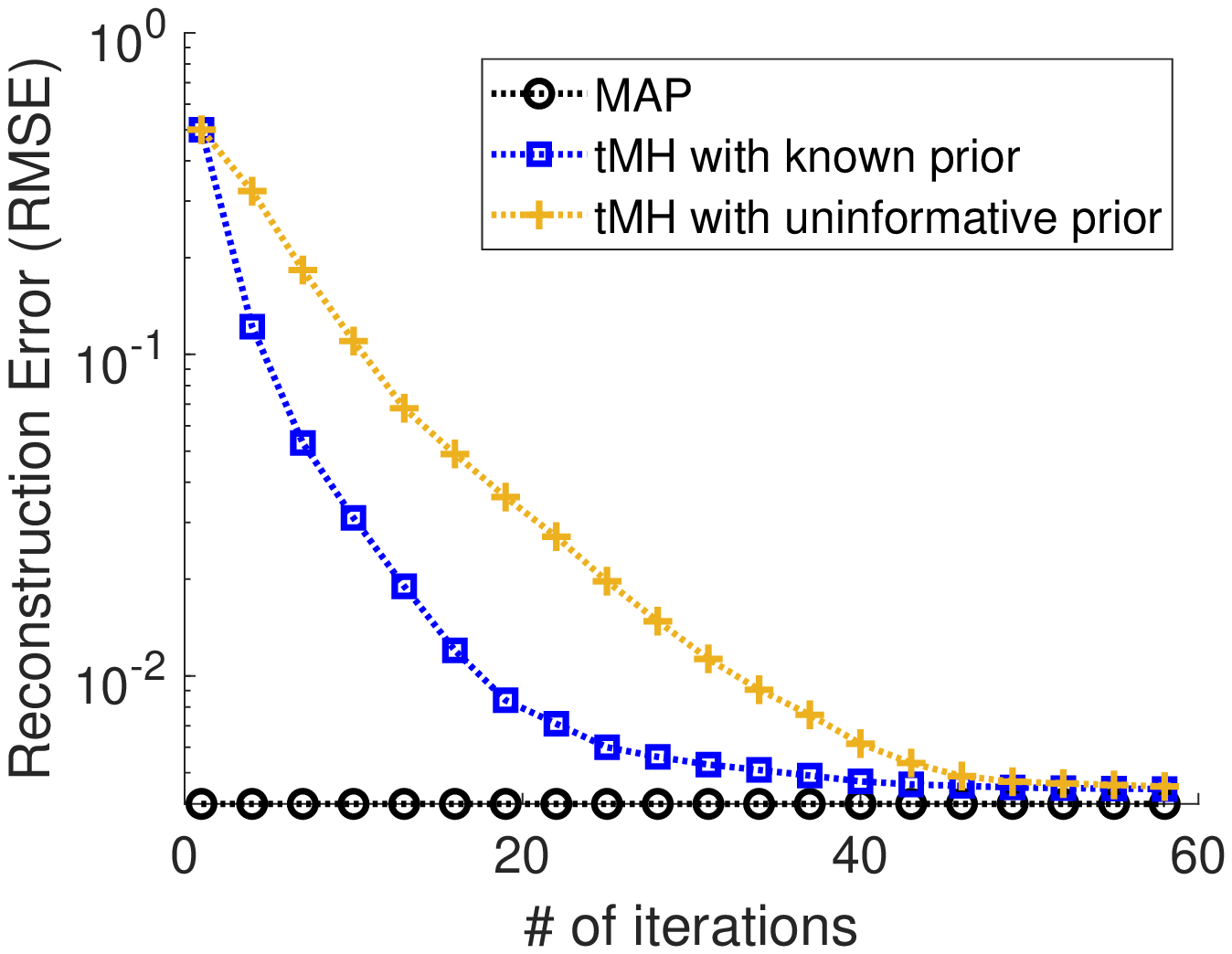}\label{fig:sim_2_3}}
\caption{(a) Comparison between the target (posterior) and the empirical distribution of a randomly chosen entry of $\mathbf{X}$ obtained via Bayesian iCR, (b) distance between the estimator empirical distribution and the estimated target distribution with respect to the number of concepts, and (c) comparison of $\mathbf{T}$ reconstruction errors (RMSE) versus the number of iterations $K$ between MAP, tMH with known prior (i.e., Bernoulli for $\mathbf{X}$ and Dirichlet for $\boldsymbol{y}$ and $\boldsymbol{z}$) distribution, and tMH with the uninformative prior (uniform) distribution. (d), (e), and (f) correspond to experiments similar to those of (a), (b), and (c), respectively, performed using the Bayesian iLCR method instead of the Bayesian iCR.}
\label{fig:algorithm_1}
\end{figure*}

\subsection{Inverse Linearized Contextual Reasoning via Compressed Sensing}
Now, let us go back to the iCR problem. Supposing that the LCR model $\mathbf{\Phi}$ is well-trained and known to all agents, what remains is to solve an underdetermined system, i.e., an iLCR problem. To solve it, we use the tMH described in Algorithm \ref{alg:tMH}, but with a different target distribution.

With a known $\mathbf{\Phi}$, the likelihood of $\bar{\mathbf{R}}$ given $\mathbf{T}$ can be written as
\begin{align}
    p(\bar{\mathbf{R}}|\mathbf{T}) &= p(\bar{\boldsymbol{r}}|\mathbf{\Phi}\boldsymbol{t},\boldsymbol{t})\\
    &= p(\boldsymbol{n}|\mathbf{\Phi}\boldsymbol{t},\boldsymbol{t})\\
    &= p(\boldsymbol{n})\\
    &\propto \exp\left(-\frac{1}{2\sigma^2}\|\boldsymbol{n}\|_2^2\right).
\end{align}
From the Bayes' rule, the posterior distribution of $\mathbf{T}$ given $\bar{\mathbf{R}}$ is
\begin{align}\label{eq:posterior_3}
    p(\mathbf{T}|\bar{\mathbf{R}}) = \frac{1}{U'}\exp\left(-\frac{1}{2\sigma^2}\| \bar{\boldsymbol{r}} - \mathbf{\Phi}\boldsymbol{t} \|_2^2\right)p(\boldsymbol{t}),
\end{align}
where $U' = \int \exp\left(-\frac{1}{2\sigma^2}\| \bar{\boldsymbol{r}} - \mathbf{\Phi}\boldsymbol{t} \|_2^2\right)p(\boldsymbol{t}) d\boldsymbol{t}$. Regarding \eqref{eq:posterior_3} as a target distribution, $\mathbf{T}$ (or equivalently $\boldsymbol{t}$) can be sampled by using the tMH. Since the first stage of the tMH finds the sparse pattern of $\mathbf{X}$ and the second stage finds the prior distributions $\boldsymbol{y}$ and $\boldsymbol{z}$, the overall process can be seen as solving a compressed sensing problem. In fact, the problem at hand can be solved using the existing compressed sensing method of the orthogonal matching pursuit (OMP) family \cite{Pati1993}. However, in order to maintain consistency with the previous Bayesian iCR method, tMH is used. The existing OMP family follows a process of finding the entry of an unknown vector that is most likely to be non-zero in the first iteration, and then repeating the process of finding the entry of the residual vectors that are most likely to be non-zero in the subsequent iterations. In contrast, the proposed method updates every unknown vector entry in every iteration. Although the proposed method requires greater computational effort, it has the advantage of being robust to error propagation.


\section{Numerical Results}\label{Sec:Simulations}
This section aims to validate the performance of the proposed Bayesian iCR and iLCR methods through numerical simulations. To validate the Bayesian iCR, we assumed that the agents utilize the original CR model for SNC. On the other hand, for validating the Bayesian iLCR, the agents utilize the trained LCR model. We assume that the random variables $\boldsymbol{y} = p(A)$ and $\boldsymbol{z} = p(C)$ follow the symmetric Dirichlet distribution of order $|\mathbb{A}|$ and $|\mathbb{C}|$, respectively, with parameters $\delta_A$ and $\delta_C$, respectively. It is worth noting that the probability density function of the symmetric Dirichlet distribution of order $M \geq 2$ with a parameter $\delta$ with respect to the Lebesgue measure on the Euclidean space $\mathbb{R}^{M-1}$ is given by
\begin{align}
    f( p_1, p_2, \dots, p_M ;\delta) = \frac{\Gamma(\delta M)}{\Gamma(\delta)^M} \prod_{m = 1}^M p_m,
\end{align}
where $\{p_m\}_{m = 1}^M \in \mathbb{P}_{M-1}$. Further we supposed that the entries of $\mathbf{X}$ are independently, identically and $\mathrm{Bernoulli}(s)$ distributed, i.e., $p(x_{c,a} = 1) = s$ and $p(x_{c,a} = 0) = 1 - s$, $\forall (c,a) \in \mathbb{C}\times\mathbb{A}$, for some real value $0 < s < 1$. Thus, $\mathbf{X}$ is generated based on the distribution. However, if $\mathbf{X}$ is generated to have more than two actions have the same set of relevant concepts, such actions are regenerated to make the matrix $\mathbf{X}$ full rank. The parameters $\theta_s$ and $\theta_r$ in \eqref{eq:RSA_recursion} are set as $\theta_s = \theta_r = 1.1$. The impact of the parameters on the performance of SNC was shown in detail through experiments in \cite{Seo2023}. Moreover the utility functions in \eqref{eq:RSA_recursion} are defined as $u_s(r_{c,a}) = \log(r_{c,a}p(c))$ and $u_r(s_{c,a}) = \log(s_{c,a}p(a))$.

\begin{figure}[t]
\centering
\subfigure[]{\includegraphics[width=0.485\textwidth]{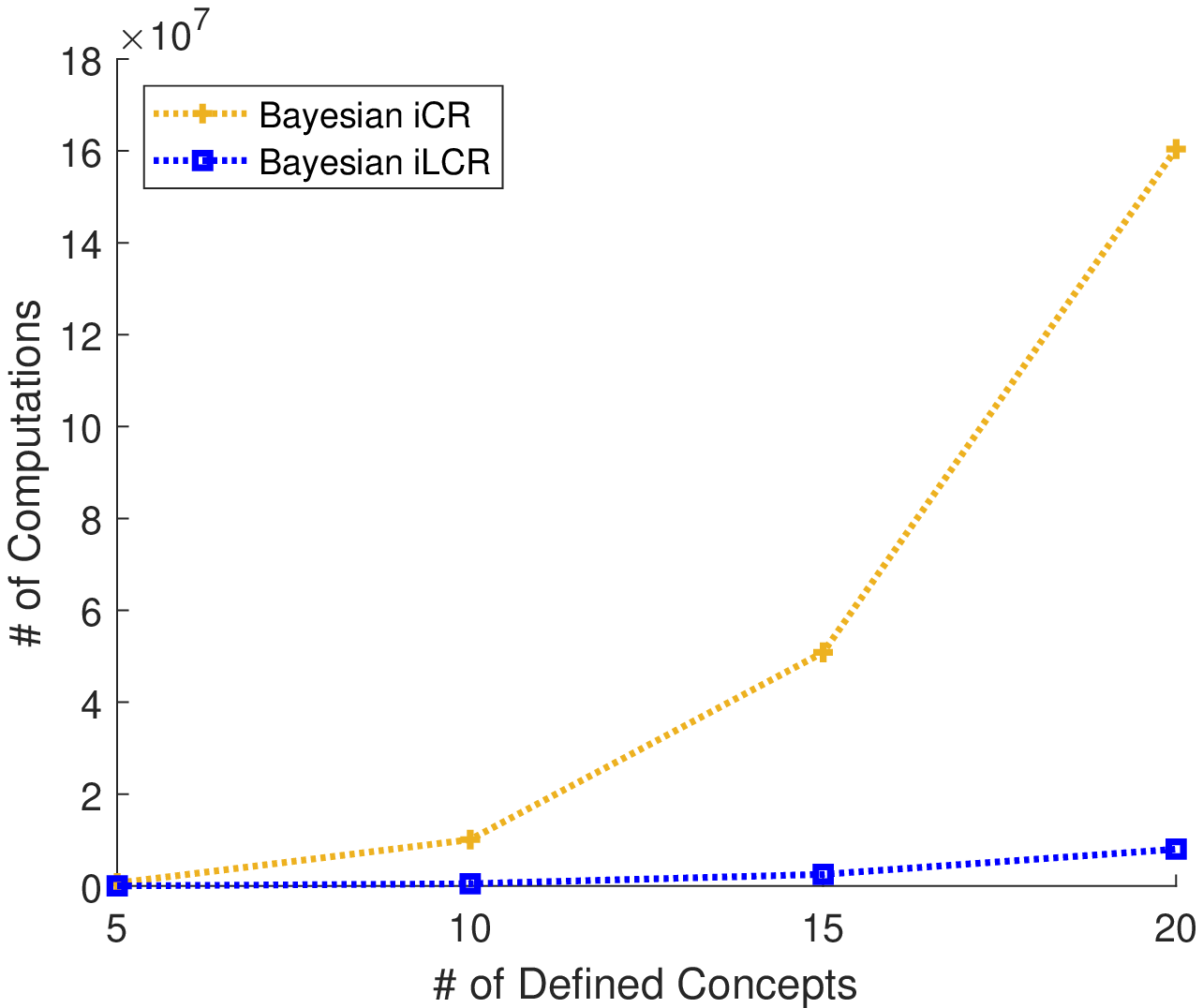}\label{fig:complexity_1}}
\subfigure[]{\includegraphics[width=0.485\textwidth]{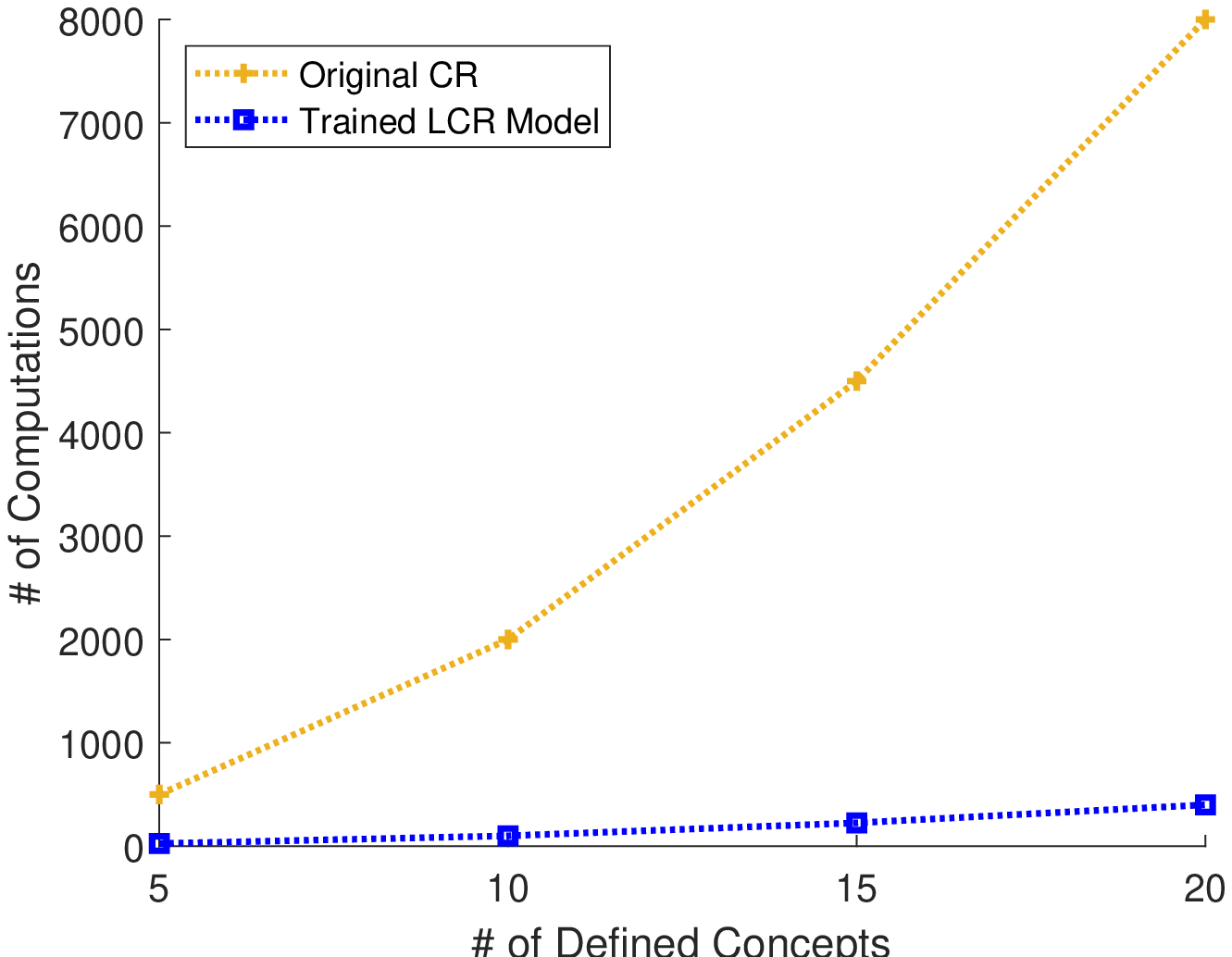}\label{fig:complexity_2}}
\caption{(a) Comparison of computational complexity between Bayesian iCR and Bayesian iLCR, and (b) between conventional CR and trained LCR model with respect to the number of defined concepts $|\mathbb{C}|$.}
\label{fig:complexity}
\end{figure}

\subsubsection{Inference Accuracy}
The inference accuracy of the Bayesian iCR and iLCR methods for solving an iCR problem is compared in Fig. \ref{fig:algorithm_1}. First, Figs. \ref{fig:sim_1_1}, \ref{fig:sim_1_2}, and \ref{fig:sim_1_3} show the performance of the Bayesian iCR method based on Algorithm \ref{alg:tMH} introduced in Section \ref{sec:BiCR}. Specifically, Fig. \ref{fig:sim_1_1} shows the target and empirical distribution of an arbitrarily chosen entry of $\mathbf{X}$ when $s = 0.3$. As shown in the figure, the two distributions exhibit a small amount of difference, however, they have nearly similar shapes. Fig. \ref{fig:sim_1_2} shows the distance between the estimator empirical distribution and the estimated target distribution with respect to the number of concepts $|\mathbb{C}|$. The distance is measured using the Jensen-Shannon divergence and averaging it over all matrix/vector entries, assuming that $|\mathbb{C}| = |\mathbb{A}|$. The distance is measured separately for each $\mathbf{X}$, $\boldsymbol{y}$ and $\boldsymbol{z}$. In addition, Fig. \ref{fig:sim_1_3} compares the $\mathbf{T}$ reconstruction error (RMSE) between MAP, tMH with known and uninformative prior (uniform prior) with respect to the number of tMH iterations $K$, with fixed $K_1 = 10$ and $K_2 = 10$. Here, because the MAP was obtained through independent operations regardless of $K$, it is represented as a constant in the figure. To obtain the numerical result of MAP, we assumed that the prior distribution is known, however, it is impractical to implement MAP since the prior distribution is unknown. From the figures, we corroborate that the Bayesian iCR method based on tMH provides relatively strong performance in terms of inference accuracy.

On the other hand, Figs. \ref{fig:sim_2_1}, \ref{fig:sim_2_2}, and \ref{fig:sim_2_3} demonstrate the performance of the Bayesian iLCR method based on the tMH by presenting the results of experiments conducted in the same format as those in Figs. \ref{fig:sim_1_1}, \ref{fig:sim_1_2}, and \ref{fig:sim_1_3}, respectively. Comparing the experimental results, it can be observed that the Bayesian iLCR method is capable of providing relatively more accurate inference. For example, when comparing Figs. \ref{fig:sim_2_1} and \ref{fig:sim_2_2} with Figs. \ref{fig:sim_1_1} and \ref{fig:sim_1_2}, respectively, it is evident that Bayesian iLCR provides an empirical distribution that is closer to the target distribution than Bayesian iCR. Similarly, when comparing Fig. \ref{fig:sim_2_3} with Fig. \ref{fig:sim_1_3}, it can be observed that Bayesian iLCR exhibits less reconstruction error.

The reason for the better performance of Bayesian iLCR over Bayesian iCR can be attributed to the difference in invertibility between the two models. Specifically, the LCR model offers better invertibility than the CR model since it can be seen as a one-to-one function as mentioned earlier. Furthermore, a loss function for training the LCR model was intentionally designed to include a term related to invertibility, which encourages the possibility of iCR. Therefore, Bayesian iLCR based on the LCR model exhibits higher inference accuracy compared to the Bayesian iCR.

\begin{figure}[t]
\centering
\subfigure[]{\includegraphics[width=0.485\textwidth]{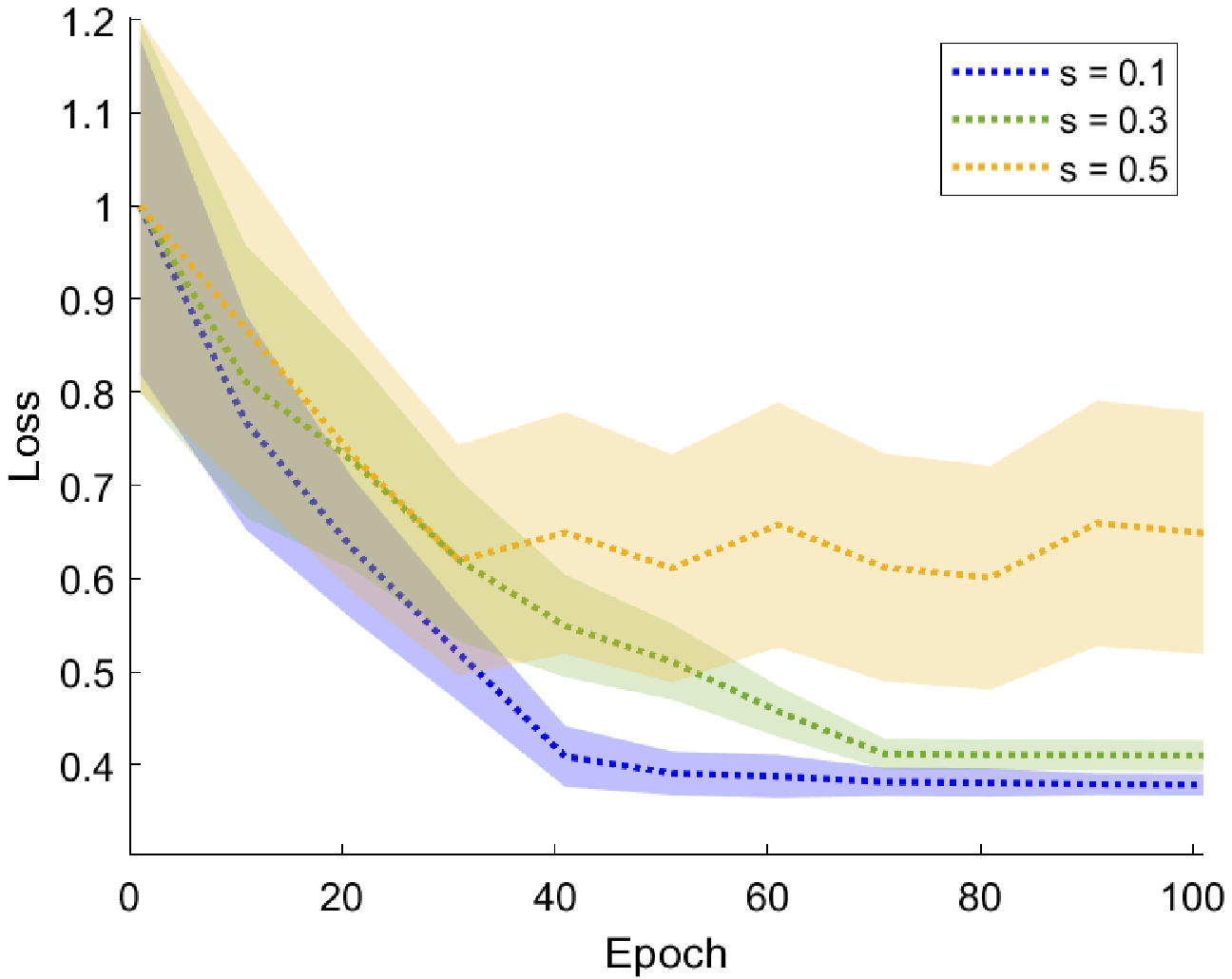}\label{fig:linearizability}}
\subfigure[]{\includegraphics[width=0.485\textwidth]{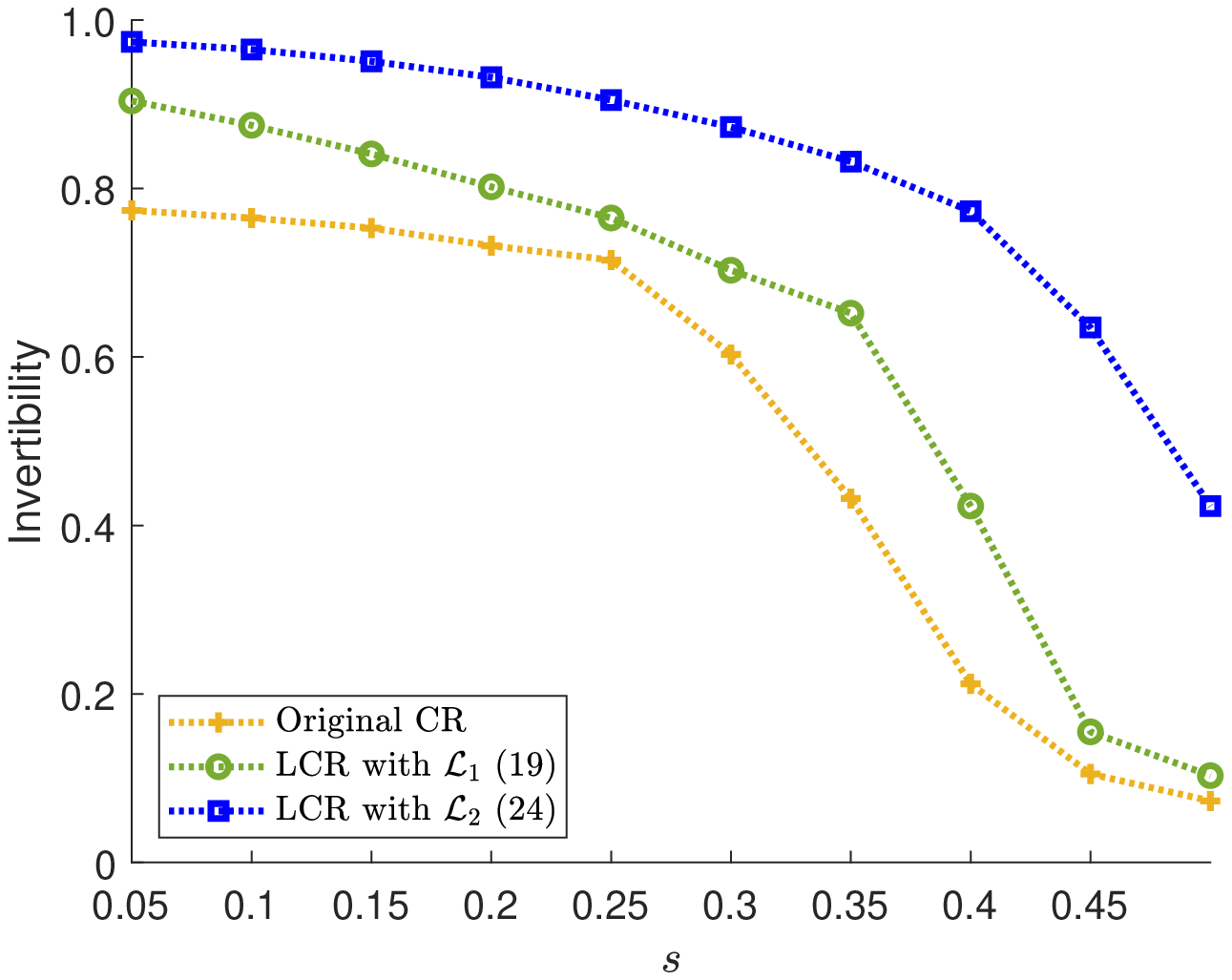}\label{fig:invertibility}}
\caption{(a) Learning curves of LCR model with respect to context sparsity $s$ and (b) invertibility of CR, LCR with different loss functions $\mathcal{L}_1$ and $\mathcal{L}_2$. Invertibility is measured by the success percentage of solving the iCR problem.}
\end{figure}

\subsubsection{Computational Complexity of Inference}
As mentioned earlier, the most significant advantage of the Bayesian iLCR method is that it greatly reduces the computational cost by using the LCR model instead of the traditional recursive method. Fig. \ref{fig:complexity_1} compares the number of arithmetic operations required to perform Bayesian iCR and Bayesian iLCR methods, respectively, with respect to the number of defined concepts $|\mathbb{C}|$ (or actions $|\mathbb{A}|$). As clearly shown in the figure, it can be observed that Bayesian iLCR requires significantly fewer arithmetic operations compared to Bayesian iCR. This is due to the difference in computational requirements between the traditional CR and the trained LCR, as demonstrated in Fig. \ref{fig:complexity_2}.

\subsubsection{Linearizability and Invertibility of CR}
As mentioned earlier, the linearizability of CR depends on context sparsity. Fig. \ref{fig:linearizability} shows the learning curve of the LCR model depending on context sparsity, where we can see that our intuition is correct. The LCR model can learn faster when the non-zero entries of the context are fewer, that is when the value of $s$ is smaller.

Based on a similar intuition, it can be understood that the invertibility of CR and LCR is also dependent on context sparsity. This intuition is particularly evident in the fact that the inverse problem of LCR can be expressed as a compressed sensing problem. Fig. \ref{fig:invertibility} shows the statistics on whether Bayesian iCR and Bayesian iLCR methods can infer the context and two prior distributions. Let's define invertibility as a percentage of success in solving the iCR problem. In particular, for the Bayesian iLCR method, invertibility was tested for the LCR model learned through two different loss functions, i.e., $\mathcal{L}_{1}$ and $\mathcal{L}_{2}$. Note that $\mathcal{L}_{2}$ is a function that adds a loss term related to the RIP condition to $\mathcal{L}_{1}$. From the figure, it is shown that the loss function design involving the quasi-RIP condition results in better invertibility of the LCR model. Furthermore, it can be noted that the linearization of the CR model itself leads to better invertibility.

\begin{figure}[t]
\centering
\includegraphics[width=0.485\textwidth]{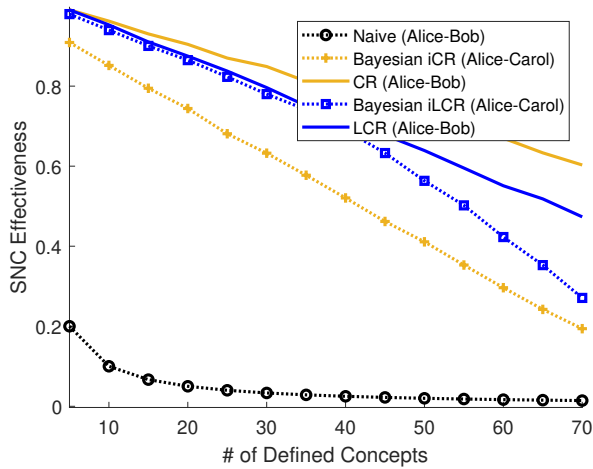}
\caption{Effectiveness of SNC between Alice and Carol based on Bayesian iCR and Bayesian iLCR, with respect to the number of defined concepts $|\mathbb{C}|$ under communication-limited environment. They are also compared with SNC between Alice and Bob with naive semantic coding, CR and LCR models.}
\label{fig:SNC_exp}
\end{figure}

\subsubsection{Heterogeneous SNC Effectiveness}
Consider the heterogeneous SNC problem involving Alice, Bob, and Carol mentioned earlier. In the following, we present experimental results demonstrating the effectiveness of context inference and SNC performed by Carol with Alice (or Bob) using Bayesian iCR and Bayesian iLCR, respectively. Additionally, we provide insights by examining the effect achieved when Alice and Bob utilize the original CR and LCR models, respectively, for SNC. Effectiveness is quantified as the probability that the receiver correctly interprets the sender's intended action through SNC.

From Fig. \ref{fig:SNC_exp}, it can be observed that the Bayesian iLCR method yields greater effectiveness when Carol communicates with Alice, compared to the Bayesian iCR method. It is worth noting that obtaining the LCR model requires a pre-training process, making a perfectly fair comparison unattainable. However, considering that CR enhances invertibility when linearized, we can observe an improvement in SNC performance in terms of effectiveness, as illustrated in Fig. \ref{fig:SNC_exp}.

On the other hand, when comparing SNC between Alice and Bob using CR and that between Alice and Carol based on iCR, a significant difference in effectiveness can be observed. This difference arises from the inability of iCR to infer a sufficiently reliable context. In contrast, when examining SNC between Alice and Bob using LCR and that between Alice and Carol based on iLCR, the difference in effectiveness is not significant. This is because, as mentioned earlier, the LCR model enhances invertibility, allowing iLCR to infer context effectively.

\begin{figure}[t]
\centering
\subfigure[]{\includegraphics[width=0.485\textwidth]{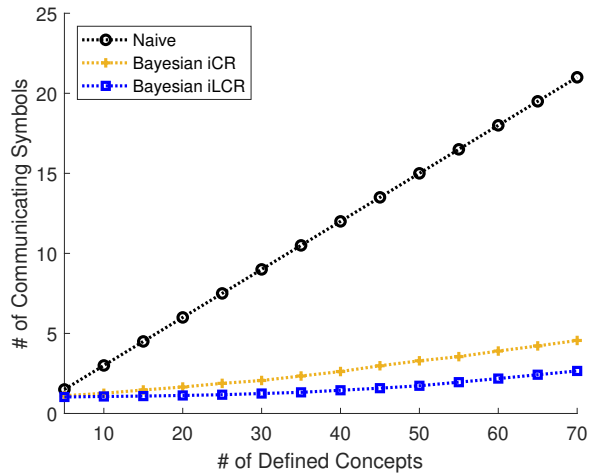}\label{fig:symbols}}
\subfigure[]{\includegraphics[width=0.485\textwidth]{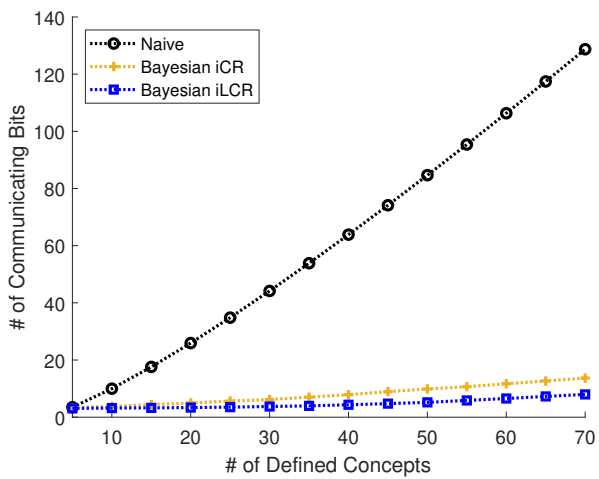}\label{fig:bits}}
\caption{(a) The number of communicating symbols and (b) bits to achieve effective SNC.}
\end{figure}

\subsubsection{Heterogeneous SNC Efficiency}
Worth noting is that the reason for the decreased effectiveness of Bayesian iCR is that it considers a situation where only one symbol is communicated at a time (a symbol per transmission), taking into account the limited communication environment. Naturally, the performance of Bayesian iCR will improve when more than one symbol can be communicated. On the other hand, Bayesian iLCR shows robust communication performance even in a limited communication environment. This is because, as explained earlier, the CR invertibility is improved during the linearization process for obtaining LCR.

Fig. \ref{fig:symbols} shows the average number of communicating symbols (symbolized concepts) that are required to achieve SNC effectiveness greater than $0.9$ under the same set of Fig. \ref{fig:SNC_exp} but in a communication-free environment, where the agents can communicate more than one symbol for SNC. Here, we compared the method of using na\"iv semantic coding, which was mentioned earlier, together with Bayesian iCR and Bayesian iLCR. Of course, if Alice and Carol use na\"iv semantic coding, they need to exchange many more symbolized concepts. On the other hand, we can see that even with a much smaller number of symbolized concepts, Bayesian iCR and Bayesian iLCR can perform effective SNC.

As was done in \cite{Seo2023}, if we consider symbolized concepts as information sources in traditional communication, we can use binary uniquely decodable source coding to determine the minimum bit-length of each symbol. Fig. \ref{fig:bits} shows the average number of bits that must be transmitted for effective heterogeneous SNC, assuming no randomness in the channel and noiseless source coding. The trends of the curves for each method shown in Fig. \ref{fig:bits} are similar to those in Fig. \ref{fig:symbols}. While na\"ive semantic coding requires a large number of bits for communication, effective SNC can be achieved with relatively few bits through context inference using Bayesian iCR and Bayesian iLCR. Therefore, the context-based heterogeneous SNC obtained through the proposed iCR solution is proven to be efficient.

\section{Conclusion}\label{Sec: Conlcusion}
In this article, we addressed the iCR problem in a heterogeneous SNC scenario using two different approaches. Firstly, we formulated the iCR problem as a Bayesian inference problem and proposed a Bayesian iCR method that utilizes the tMH algorithm. The proposed method guarantees a certain level of inference performance, yet comes with a considerable computational cost. To obviate this, we linearized the CR using a neural network to obtain the LCR model and replace the iCR problem with a compressed sensing problem. Similar to the previously proposed method, we further proposed the Bayesian iLCR method to solve this compressed sensing problem. As a result, we solved the iCR problem with minimal computational cost and enabled SNC between agents with different contexts.

\bibliographystyle{IEEEtran}
\bibliography{References}

\end{document}